 \documentclass[final,5p,times,twocolumn,authoryear]{elsarticle}

\usepackage{amssymb}
\usepackage{amsmath}
\usepackage{newtxtext}
\usepackage[varvw]{newtxmath}
\usepackage{natbib}
\usepackage{hyperref}

\newcommand{\msun}{$\,M_\odot$}
\newcommand{\Msun}{\,M_\odot}
\newcommand{\rsun}{$\,R_\odot$}

\def\orb{\mathrm{orb}}
\def\oorb{\Omega_\mathrm{orb}}
\def\crit{{\mathrm{crit}}}
\def\eff{{\mathrm{eff}}}
\def\gw{{\mathrm{gw}}}
\def\mb{{\mathrm{mb}}}
\def\mass{{\mathrm{mass}}}

\def\BHL{{\mathrm{BHL}}}
\def\Mach{{\mathcal{M}}}
\def\foc{{\mathrm{foc}}}
\def\sync{{\mathrm{sync}}}
\def\circ{{\mathrm{circ}}}
\def\tide{{\mathrm{tide}}}

\def\disc{{\mathrm{disc}}}
\def\tot{{\mathrm{tot}}}
\def\res{{\mathrm{res}}}

\def\ini{{\mathrm{ini}}}
\def\donor{{1}}
\def\gainer{{2}}
\def\Roche{{\mathrm{Roche}}}
\def\RL{{$\mathcal{R}_L$}}
\def\loss{{\mathrm{loss}}}
\def\wind{{\mathrm{wind}}}
\def\RLOF{{\mathrm{RLOF}}}

\def\Mtrans{{\dot{M}_{1\leftrightarrow 2}}}

\def\rot{{\mathrm{rot}}}
\def\bin{{\mathrm{bin}}}
\def\spin{{\mathrm{spin}}}

\def\thin{{\mathrm{thin}}}
\def\thick{{\mathrm{thick}}}

\def\cond{{\mathrm{cond}}}
\def\eff{{\mathrm{eff}}}
\def\rem{{\mathrm{re-emi}}}
\def\jeans{{\mathrm{Jeans}}}
\def\impact{{\mathrm{impact}}}
\def\iso{{\mathrm{iso}}}
\def\acc{{\mathrm{acc}}}
\def\Edd{{\mathrm{Edd}}}

\def\phot{{\mathrm{phot}}}
\def\myr{$\,M_\odot\,$yr$^{-1}$}

\def\lone{{\mathcal{L}_1}}
\def\Lone{${\mathcal{L}_1}$}
\def\Ltwo{${\mathcal{L}}_2$}
\def\Lthree{${\mathcal{L}}_3$}
\def\LL{\mathcal{L}}

\journal{New Astronomy}

\begin{document}

\begin{frontmatter}



\title{Evolving Low- and Intermediate-Mass Binaries: Departures from Classical Theory}

\author[first]{Lionel Siess}
\ead{lionel.siess@ulb.be}
\ead[url]{http://www.astro.ulb.ac.be/~siess/}
\affiliation[first]{organization={Institut d'Astronomie et d'Astrophysique and BLU-ULB (Brussels Laboratory of the Universe), Université Libre de Bruxelles (ULB)},
            addressline={Avenue F.D. Roosevelt 50, CP226}, 
            city={Brussels},
            postcode={1050}, 
            country={Belgium}}

\begin{abstract}
This review explores the physical mechanisms driving the evolution of low- and intermediate-mass binary star systems, with particular emphasis on emerging mechanisms that challenge classical paradigms. We begin by describing the principal formation channels and orbital properties of binary systems. A critical reassessment of the Roche lobe formalism is presented, focusing on systems with eccentric orbits and asynchronous rotation, where deviations from traditional approximations become significant.  We then review current theoretical models of mass and angular momentum exchange via Roche-lobe overflow, incorporating results from recent hydrodynamical simulations of wind accretion. The review also reports advances in tidal dissipation theory. Finally, we explore mechanisms capable of sustaining or exciting orbital eccentricity, including perturbations induced by mass transfer and interactions with circumbinary disks. These discussions aim to outline underexplored facets of binary evolution, offering new perspectives for theoretical and observational studies.

\end{abstract}




\end{frontmatter}




\section{Introduction}
\label{introduction}

Binary stars are central to astrophysics, providing the framework for understanding diverse stellar phenomena and serving as key actors in cosmic evolution. They provide evolutionary pathways to the formation of exotic objects such as blue stragglers, novae, Type Ia supernovae (SNIa), cataclysmic variables or gamma ray-bursts. They are also invoked to explain chemically peculiar stars such as Ba or CH stars or in the shaping of circumstellar environment of planetary nebulae and evolved stars.

Binary stars also play a key role in galactic chemical enrichment. Material ejected during their lifetimes, as well as ejecta from their potentially explosive endpoints, contributes significantly to the enrichment of the interstellar medium. For instance, the decompression of neutron-rich matter during neutron star mergers has emerged as a promising site for the nucleosynthesis of rapid neutron capture (r-process) elements \citep[e.g.][]{Just2023}, while iron is known to be produced mainly by SNIa \citep{Matteucci1986}.

Binary stars are also sources of gravitational wave, produced by the merging of compact binaries hosting black holes and neutron stars. These events provide direct observational evidence of binary interactions and open a new window into the study of compact objects, offering valuable insights into their properties and the fundamental nature of gravity.
In the context of planet formation, binaries exert a major influence: their dynamical environments shape the formation and subsequent evolution of planetary systems, leaving an imprint on the distribution and properties of exoplanets. 
Moreover, binary stars enable precise determinations of stellar masses and, in some cases, of their radius. These measurements are fundamental for testing stellar models and understanding stellar structure and evolution. Observations of binary systems also provide constraints on binary physics, helping to refine models of stellar interactions, jet formation, accretion disk dynamics, and evolutionary pathways. Through SNIa, binaries also provide a means to estimate galactic distances, making them crucial tools in cosmology. Their influence extends across many areas of astrophysics, not only in stellar evolution, nucleosynthesis, and gravitational wave astronomy, but also as natural laboratories for studying complex physical processes associated with the exchange of mass and angular momentum. As such, they offer unique opportunities to probe and refine our understanding of fundamental astrophysical processes.

This paper does not aim to give a full review of binary star evolution, a topic already well covered in recent textbooks and review articles \citep[e.g.][]{Eggleton_book,Tauris_vdH2023,Chen2024,Marchant2024}. Instead, it offers a concise overview of the key physical processes included in binary evolution models, with a particular emphasis on several aspects that have received comparatively less attention. The discussion focuses primarily on low- and intermediate-mass binary systems, which serve as the source for many of the illustrative examples presented here.

Before outlining the structure of the review, it is helpful to briefly situate these physical processes within the observational landscape. Roche-lobe overflow, wind accretion, tidal coupling, and circumbinary disk torques manifest in a variety of observed systems.  Classical Algol systems exemplify stable, donor-to-accretor mass transfer; barium stars and symbiotic binaries illustrate the importance of wind-driven accretion; and tidal interactions leave clear imprints in the light curves and rotational states of ellipsoidal variables and close double stars. These well-studied observational classes provide natural anchor points that link theoretical formalisms to empirical phenomena. By framing the underlying mechanisms in this way, the revised prescriptions presented in this review aim to establish a clearer conceptual bridge between physical modeling and the diverse behaviors exhibited by interacting binaries.

The structure of this review is as follows. Section \ref{sect:formation} offers a brief overview of binary star formation, followed by an analysis of their orbital properties in Section \ref{sect:stats}. Section \ref{sect:physics} delves into the physics of binary interactions, beginning with a critical reassessment of the Roche lobe concept in eccentric and asynchronously rotating systems (Section \ref{sect:roche}). This is followed by a review of current models for mass and angular momentum transfer via Roche-lobe overflow (Section \ref{sect:RLOF}). We then examine the influence of wind-driven mass loss and accretion on orbital dynamics, with particular attention to recent hydrodynamical simulations of wind Roche-lobe overflow (Sections \ref{sect:wind_loss}–\ref{sect:wind_acc}). Prescriptions for systemic mass and angular momentum loss are discussed in Section \ref{sect:systemic}, while theories of tidal interactions are presented in Section \ref{sect:tides}. Mechanisms responsible for eccentricity excitation, whether driven by mass transfer or interactions with circumbinary disks, are explored in Sections \ref{sect:eccentric} and \ref{sect:CBD}. The review concludes with a synthesis of these insights and a discussion of future research directions.

\section{A short history of binary formation}
\label{sect:formation}

Given their far-reaching impact, understanding how binary stars form is a central question in stellar astrophysics. Binary formation theories must account for the wide diversity of observed systems, ranging from tight, interacting binaries to wide, detached pairs, and reconcile this with the physical conditions of star-forming regions. The ideas behind the formation of binary stars have not evolved much in the past decades, but progress has been made thanks to the enhanced computational capabilities of modern computers, and some more likely scenarios have emerged.  
Three main processes are generally invoked in the formation of binary stars \citep[e.g.][]{Pringle1991,Tohline2002,Kratter2011}. The oldest proposition is the fission scenario, in which a rotating gas cloud contracts and as a result of angular momentum conservation spins up. The core then evolves through a succession of ellipsoidal figures of equilibrium. With increasing rotational speed, triaxial deformations develop, producing a flattening and an elongation of the core that becomes oblate. Eventually, the structure bifurcates into pear and dumbbell shapes, leading to fission and the formation of two detached rotating masses. This model assumes that the core is rotating uniformly, an unrealistic simplification that has been largely dismissed by hydrodynamical simulations \citep[e.g.][]{Durisen1986}.

An alternative scenario is the capture model, in which two isolated stars come sufficiently close to each other and dissipate enough kinetic energy to remain gravitationally bound. The energy dissipation process can be attributed to tidal forces \citep{Fabian1975}, dynamical friction and drag within the circumstellar environment of protostars \citep{Silk1978,Rozner2023}, interactions with an accretion disk \citep{Clarke1991} or three-body dynamical encounters where the excess energy can be transferred to the third body \citep[e.g.][]{Atallah2024}. The capture scenario is favored in dense stellar environments like in globular clusters, but is generally thought to play a minor role in the overall formation of binary systems.

The most promising theories for binary formation involve fragmentation processes. The mechanism often referred to as prompt or early fragmentation \citep{Hoyle1953,Inutsuka1992,Bodenheimer2000} considers fragmentation of the protostellar cloud. During the gravitational collapse of a self-gravitating, rotating and likely turbulent molecular cloud, gravitational instabilities develop in the overdensities where the Jeans mass is exceeded. These regions collapse on a shorter timescale than the overall cloud, causing its fragmentation into smaller clumps or filaments. If the fragments are sufficiently close and massive, they can form gravitationally bound pairs, resulting in the creation of binary or multiple star systems. The outcome of fragmentation is highly dependent on the thermal and dynamical evolution of the cloud, with processes like turbulence, cooling, and feedback from radiation influencing the final binary system's properties. 
Hydrodynamical simulations \citep{Bate2012} have shown that fragmentation can result in a broad spectrum of binary properties, such as orbital separation, mass ratios, eccentricity, properties that depend on the initial conditions in the molecular cloud (e.g., angular momentum, initial density, and magnetic field strength). The alternative or complementary scenario is disk fragmentation that occurs in a rotating protostellar disk \citep{Bonnell1994,Kratter2016}. As for the fission scenario, the disk can become gravitationally unstable and, if cooling is sufficiently efficient, it can fragment to form small cores. These fragments can then grow by accreting mass from the surrounding material. Continued mass infall and efficient cooling are critical conditions for disk fragmentation to occur \citep{Bate_Bonnel1997, Kratter_etal2011}. This mechanism typically leads to the formation of close binary systems \citep{Bate_Bonnel1997,Kratter_etal2011}. 

\section{Orbital properties of binary stars}
\label{sect:stats}

Thanks to increasing instrumental sensitivity and the growing volume of data from large-scale surveys, our understanding of binary star demographics is becoming increasingly detailed \cite[for extensive reviews, see][]{Duchene2013,Moe2017}. 
Observations indicate that the binary fraction rises with stellar mass and is generally higher in field populations than in star clusters, where frequent dynamical interactions can disrupt binaries or inhibit their formation altogether. The binary fraction of field main sequence M-type stars increases from $\approx 0.1$ to 0.4 between 0.1 and 0.6\msun. For solar type stars it is of the order of $\approx 0.5$ and exceeds 80\% in massive stars with $M\gtrsim 10\Msun$ \citep{Sana2012}. For low-mass stars, the binary fraction is higher in younger pre-main sequence stars than in older main sequence stars \citep{Duchene2007}. Concerning the mass ratio, the distribution for high mass stars is relatively flat and approaching one which corresponds to equal-mass binaries. Solar type stars (FGK dwarfs) have a mass ratio distribution roughly flat between $q=0.1$ and 1.0 and for the lower-mass M-dwarfs the distribution is strongly peaked around $q=1$ but strong biases are present toward equal masses. The eccentricity distribution depends on the orbital period, mass, and evolutionary status of the stars. Main sequence stars with short period systems ($P \lesssim 10$) days are tidally circularized (see Sect.~\ref{sect:tides}), for intermediate periods ($10 < P \lesssim  1000$)  there is a wide range of eccentricities. 

\section{Input physics for binary evolution}
\label{sect:physics}

As stars evolve, changes in their internal structure can significantly impact the orbital dynamics of binary systems, often leading to mass and angular momentum transfer between the stellar components. Binary evolution models seek to capture how these stellar transformations modify key orbital parameters such as the semi-major axis, eccentricity, and spin. 
This section presents an updated overview of the fundamental physical processes that govern binary interactions. 
We begin by revisiting the concept of the Roche lobe, which delineates the shape of the stars and the conditions under which mass transfer between the components becomes possible. A detailed understanding of the Roche geometry enables the quantification of mass and angular‑momentum transfer through Roche-lobe overflow (RLOF), providing the foundation for modeling many interacting binaries. We then extend this framework to include wind mass loss and accretion, examining their effects on the orbital parameters and gathering insight from recent hydrodynamical simulations. Subsequent sections address the tidal interactions and the treatment of eccentric orbits in binary evolution codes. Finally, we consider the influence of circumbinary discs, whose presence may be a source of eccentricity. Together, these processes offer a cohesive picture of the complex pathways that govern binary‑system evolution.

\begin{figure}
    \centering
    \includegraphics[width=1\linewidth]{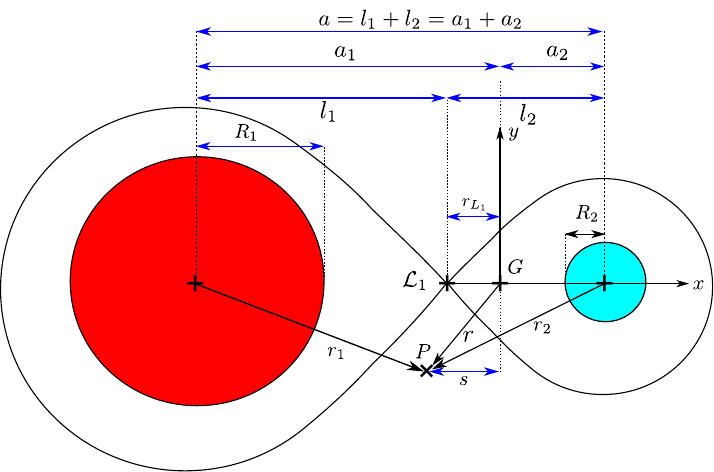}
    \caption{Notations and system geometry. $G$ is the center of mass of the binary, which consists of a primary and a secondary star with radii $R_1$ and $R_2$, respectively. The inner Lagrangian point, \Lone{}, lies along the line connecting the centers of the two stars, which defines the x-axis of the coordinate system.}
    \label{fig:notation}
\end{figure}

\subsection{Roche potential}
\label{sect:roche}

For stars in circular orbits with synchronous rotation and treated as point masses (so tidal deformations can be neglected and the Poisson equation can be solved analytically), the Roche potential at a point $P$ corotating with the binary is given by 
\begin{equation}
\displaystyle \phi_\Roche=-\frac{GM_{\donor}}{r_{\donor}}-\frac{GM_{\gainer}}{r_{\gainer}}-\frac{1}{2}\oorb^{2}s^{2}\ ,
\end{equation}
where $r_\donor$ and $r_\gainer$ are the distances from point $P$ to the centers of the donor (primary) star of mass $M_\donor$ and of the gainer (secondary/accretor) of mass $M_\gainer$. $\oorb$ is the orbital velocity and $s$ the distance of $P$ to the rotation axis passing through the center of mass $G$ (see Fig.~\ref{fig:notation}).

\begin{figure}[t]
    \centering
    \includegraphics[width=1\linewidth]{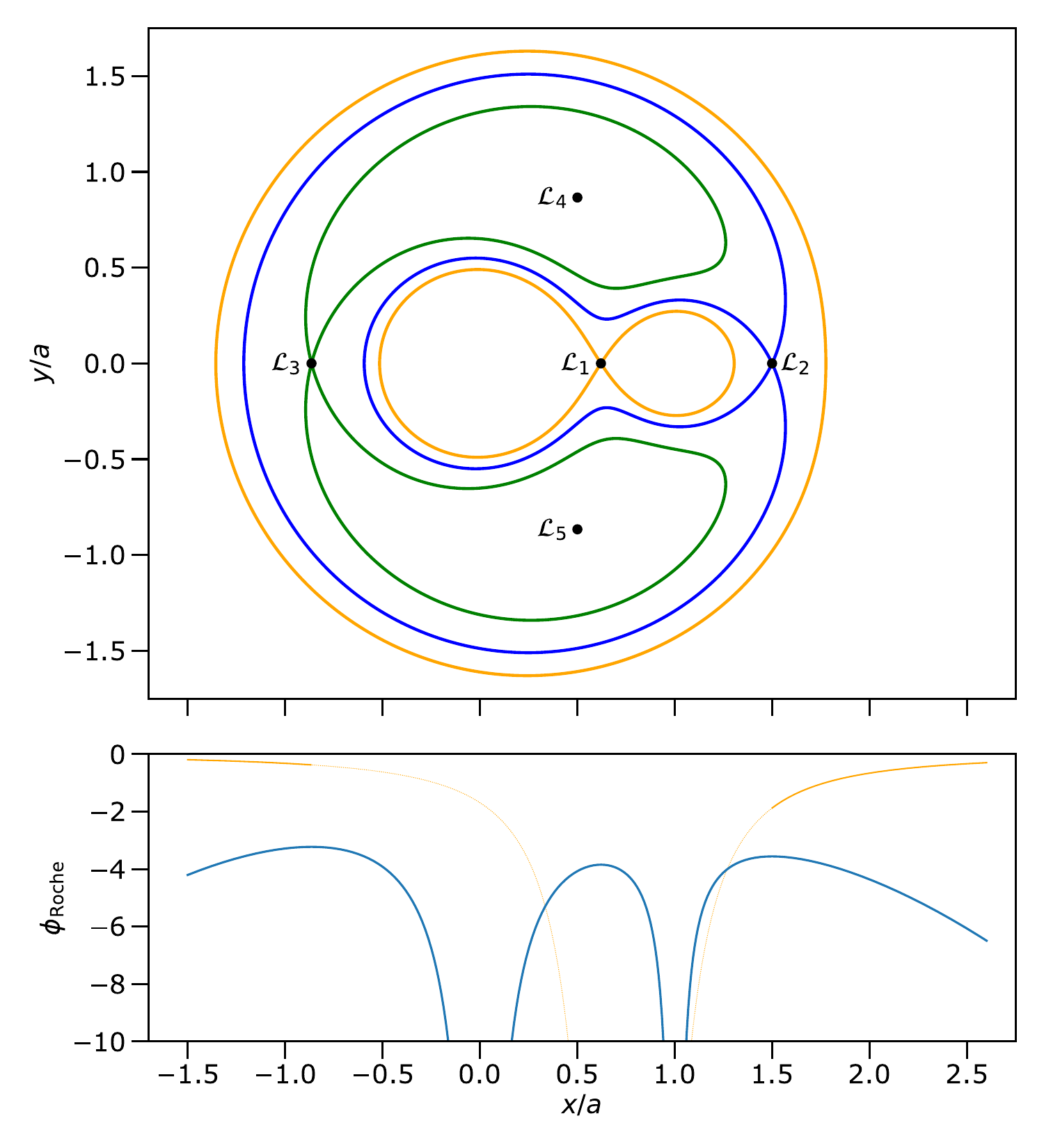}
    \caption{The top panel displays the equipotential surfaces passing through the Lagrangian points \Lone{} (orange), \Ltwo{} (blue) and \Lthree{} (green) for a mass ratio $q=M_\donor/M_\gainer=0.3$. The bottom panel shows the Roche potential along the x-axis joining the two stars (blue line). The orange line represents the binary gravitational potential at large distances ($\phi_\bin = -G(M_\donor+M_\gainer)/r$), where the system acts as a point mass for a test particle.}
    \label{fig:Equipot}
\end{figure}

In the rotating coordinate system of the binary stars, the motion of a fluid element with a velocity $v$ is governed by
\begin{equation}
    \frac{d v}{dt} +2 \,\Omega_\orb \wedge v = -\frac{\nabla P_\tot}{\rho} -\nabla \phi_\Roche\ ,
    \label{eq:motion}
\end{equation}
where $P_\tot$ is the total pressure including both gas and radiation contributions, and $\rho$ the density. When the stars are in co-rotation, the fluid element comprising the stellar material is stationary in the rotating frame ($v= 0$), and Eq.~\ref{eq:motion} reduces to
\begin{equation}
   \frac{\nabla P_\tot}{\rho} + \nabla \phi_\Roche = 0 \ .
    \label{eq:hydrostatic}
\end{equation}
This hydrostatic equilibrium condition implies that $P_\tot$ and $\rho$ are functions of the potential $\phi_\Roche$. At the stellar surface where $P_\tot$ and $\rho$ vanish, the shape of the star coincides with an equipotential surface defined by  $\phi_\Roche = \mathrm{constant}$. 
It should be emphasized that at large distances, where co-rotation can no longer be maintained, the Roche potential ceases to be valid. In this regime, it must be replaced by the gravitational potential of the entire binary system, $\phi_\Roche \approx \phi_\bin = -G(M_\donor+M_\gainer)/r$, where $r$ is the distance of the particle to the center of mass of the system (see Fig.~\ref{fig:Equipot}).

The Roche potential $\phi_\Roche$ has 5 extrema, corresponding to the Lagrangian points $\LL_{i=1,5}$ conventionally ordered by increasing potential such that $\phi_\Roche(\LL_{i})<\phi_\Roche(\LL_{i+1})$. These points are equilibrium positions where the net force on the test particle vanishes.
Figure \ref{fig:Equipot} illustrates the Roche equipotential surfaces for a binary system with a mass ratio of $q=0.3$. The equipotential surface forming the characteristic figure-eight shape and passing through the \Lone{} point delineates the limiting boundary separating the two stellar components. Material within a star's Roche lobe remains gravitationally bound to that star. The Roche lobe concept is fundamental for determining the onset of mass transfer in binary systems and for classifying their configurations as detached, semi-detached, or contact, depending on whether neither, one, or both stars fill their respective Roche lobes. 
An accurate analytical estimate of the volume-equivalent Roche radius is given by
\citep{Eggleton1983} 
\begin{equation}
R_{\lone,i}=\frac{0.49\,q_{i}^{2/3}}{0.6q_{i}^{2/3}+\ln(1+q_{i}^{1/3})}\,a\ ,
\label{eq:Roche_radius}
\end{equation}
where $a$ is the semi-major axis and $q_i$ the mass ratio defined as
\begin{equation}
q_{i}=M_{i}/M_{3-i}\ .
\label{eq:mass_ratio}
\end{equation}
The Roche lobe radius of the companion star can be directly estimated using the reverse mass ratio, i.e. $q_{3-i}=q_{i}^{-1}$. 

A strong limitation of this model is the assumption of synchronous stellar rotation, which does not necessarily hold even in circular orbits. \cite{Plavec1958} and \cite{Limber1963} first investigated the situation in which the orbital and stellar angular velocities differ in magnitude but their vectors remain parallel. Their studies, conducted independently, demonstrated that an alternative formulation of the Roche potential can be derived under the condition that the stellar interior rotates uniformly with angular velocity $\Omega_\star$. In the star's rotating frame, this condition implies that the velocity and acceleration of a mass element are negligible \citep[for a detailed analysis, see][]{Limber1963}. The resulting expression of the potential for an asynchronous donor becomes
\begin{equation}
   \displaystyle \phi_\Roche=-\frac{GM_{\donor}}{r_{\donor}}-\frac{GM_{\gainer}}{r_{\gainer}}-\frac{1}{2}\oorb^{2}s^{2} -\frac{1}{2}(\Omega_\star^2-\oorb^{2})r_\donor^{2} \ , 
   \label{eq:roche_async}
\end{equation}
where the notations are the same as before.
With this formulation of the potential, an increase in stellar spin reduces the effective gravity, causing the \Lone{} point to shift closer to the rotating star. At sufficiently high rotational velocities, \cite{Plavec1958} found that the position of \Lone{} becomes nearly insensitive to the direction of rotation, whether prograde or retrograde.
The stellar surface of the asynchronous component readjusts in response to the modified effective potential, deviating from the configuration expected under synchronous rotation. Consequently, the star establishes its own effective neutral point where the resulting gravitational and centrifugal forces balance out. This position is determined by the degree of asynchronism and the binary mass ratio. In the case of synchronous rotation, the neutral point of each star merges to become a unique \Lone{} point. This decoupling alters the topology of the effective potential, significantly affecting the dynamics of mass transfer and the evolution of angular momentum. In particular, for super-synchronous rotation ($\Omega_\star > \oorb$), mass transfer can be initiated earlier than predicted by standard, synchronous model.
Furthermore, matter at the stellar surface exhibits a non-zero velocity component in the rotating frame. When combined with the presence of tidal deformation, it becomes possible that matter escapes the star from outside the neutral point \citep{Kruszewski1963}.

\begin{figure}[h]
    \centering
    \includegraphics[width=\columnwidth]{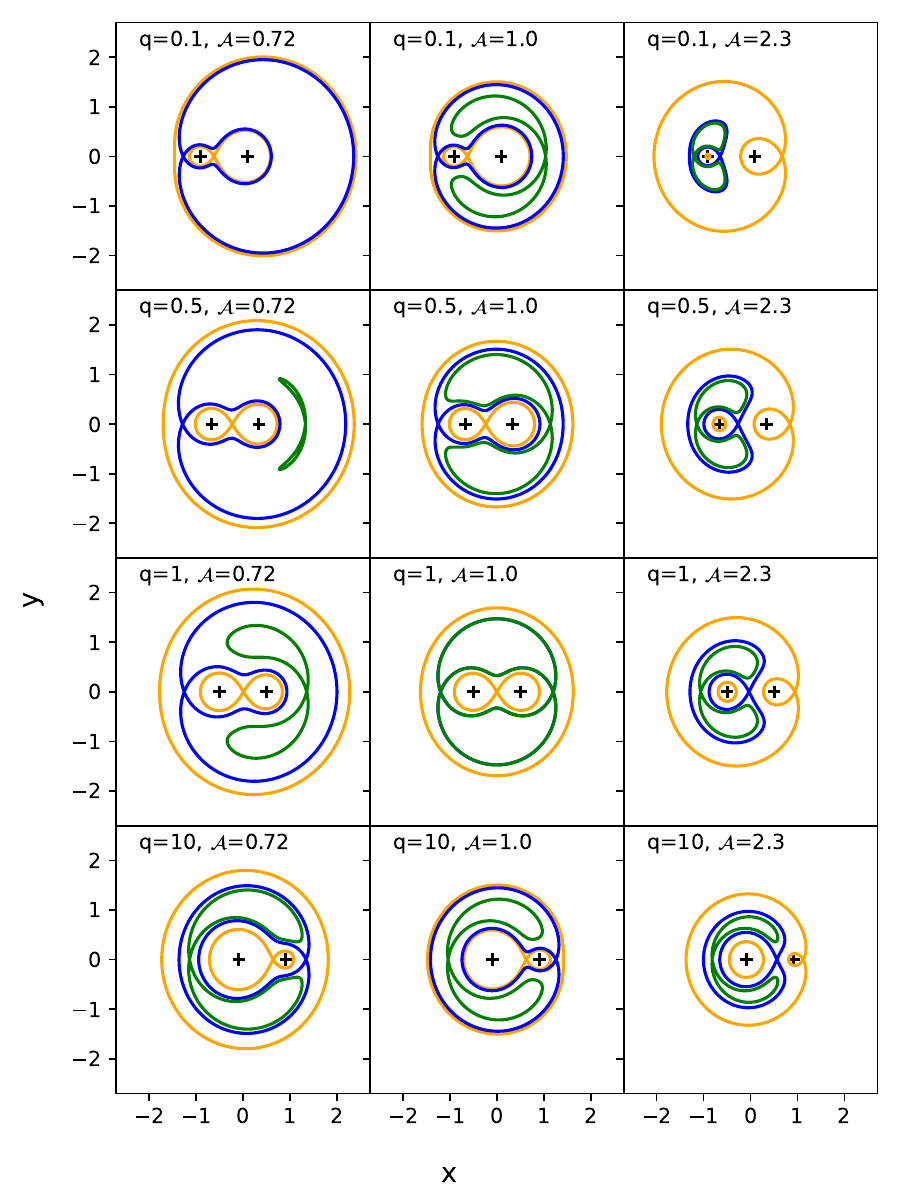}
    \caption{Topology of the Roche equipotentials in the orbital plane as a function of mass ratio $q=M_1/M_2$ and asynchronous parameter $\cal{A}$. Star 1 is located on the left side. The inner yellow equipotential passes through the \Lone{} point and the green and blue curves through \Ltwo{} and \Lthree, respectively. For $\mathcal{A} = 2.3$, one notices the absence of a critical \Lone{} equipotential enclosing the 2 stars.}
    \label{fig:roche}
\end{figure}

The dependence of the Roche potential on eccentricity was initially introduced by \cite{Avni1976} and later generalized by \cite{Sepinsky2007} to include the effects of stellar asynchronism.  The variability of the orbital separation and the relative angular velocities along the eccentric orbit introduces an explicit time dependence in the Roche potential, which drives tidal oscillations within the stellar components. \cite{Sepinsky2007} showed that if the dynamical timescale of these oscillations is much shorter than the tidal timescale defined as $\tau_\tide = 2\pi/|\oorb -\Omega_\star|$, the star can be considered at each time in quasi-static equilibrium and its shape can be determined from a generalized potential of the form  
\begin{equation}
\phi_\Roche=-\frac{GM_2}{d}\left[\frac{q}{\tilde{r}_{\donor}}-\frac{1}{\tilde{r}_{\gainer}}-\frac{1}{2}\mathcal{A}(1+q)(\tilde{x}^{2}+\tilde{y}^{2})+\tilde{x}\right],
\label{eq:phi_sepinsky}
\end{equation}
where the coordinates $(\tilde{x},\tilde{y})$ are expressed in the reference frame attached to the asynchronous star and the distances ($\tilde{r} = r/d$) are normalized to the instantaneous separation
\begin{equation}
d(\nu)=\frac{a(1-e^{2})}{1+e\cos\nu}\ ,
\label{eq:inst_sep}
\end{equation}
with $e$ is the eccentricity and $\nu$ the true anomaly.  The degree of eccentricity and asynchronism is parameterized \citep{Sepinsky2007} by 
\begin{equation}
\mathcal{A}=\frac{f^{2}(1+e)^{4}}{(1+e\cos\nu)^{3}} = f^2 \frac{1+e}{(1-e)^3}\left(\frac{d}{a}\right)^3\ ,
\label{eq:def_async}
\end{equation}
where $f$ is the spin angular velocity of the star in units of the orbital angular speed at periastron
\begin{equation}
f=\frac{\Omega_\star}{\oorb}\ ,
\label{eq:def_f}
\end{equation}
and the instantaneous orbital angular velocity
\begin{equation}
\oorb(\nu)=\frac{2\pi}{P}\frac{\left(1+e\cos\nu\right)^{2}}{\left(1-e^{2}\right)^{3/2}}\ ,
\label{eq:omega}
\end{equation}
with the orbital period $P=(4 \pi^2 a^3/(G (M_\donor+M_\gainer))^{1/2}$. Figure ~\ref{fig:roche} illustrates the effects of varying the mass ratio and asynchronous parameters $\mathcal{A}$ on the Roche geometry. As in the standard case, when the mass ratio increases, the Roche lobe around the more massive star grows larger, and the \Lone{} point shifts closer to the less massive component. For the mass ratios considered, the shape of the equipotential is sensitive to the value of $\mathcal{A}$. Remarkably, for $\mathcal{A} > 1$, the equipotential surface passing through the inner Lagrange point no longer encloses the companion star. In this configuration, the gravitational potential at the outer \Ltwo{} point can exceed or fall below that at \Lone{}, depending on the specific value of $\mathcal{A}$. This opens the possibility for a fraction of the mass flowing through \Lone{} to escape the system.  For large values of  $\mathcal{A} > 10$ or high mass ratios, the equipotential surfaces approach those expected for a single star.  With the introduction of the parameter $\mathcal{A}$, the size of the Roche lobe is also modified, but a simple solution consisting of substituting the semi-major axis $a$ with the instantaneous separation $d$ in Eq.~\ref{eq:Roche_radius} does not provide an accurate estimate of the Roche radius \RL{}  \citep{Sepinsky2007}. In fact, the resulting errors can be as large as 40\% for extreme values of $q$. 

Radiation further perturbs the Roche potential by reducing effective gravity through photon absorption and scattering by atoms \citep{Castor1975}, molecules \citep{Jorgensen1992}, or dust \citep{Bowen1988}. Conceptually, one has to distinguish if the point under consideration lies in an optically thin or optically thick region \citep{Friend1982,Dermine2009}. In the optically thick regime, radiative transfer can be described using the diffusion approximation, yielding an isotropic radiation pressure. When these layers are in hydrostatic equilibrium, the gradient of total pressure (gas+radiation) balances the gravitational force (Eq.~\ref{eq:hydrostatic}). In such cases, radiative acceleration is implicitly accounted for, and the gravitational potential remains unaltered. Conversely, in optically thin regions, the free-streaming approximation applies. Here, the radiation pressure gradient gives rise to a repulsive radiative acceleration that is radial, scales with $1/r^2$, and must be explicitly included to reduce the effective gravitational potential.
In the region close to the photosphere between the optically thick and thin regimes or in layers where the assumption of hydrostatic equilibrium breaks down, the situation is more ambiguous \citep{Dermine2009}. 

The first attempt to include the effect of radiation from one component was formulated by \cite{Schuerman1972} and simply consists of replacing the inertial mass of the star by $M(1-\Gamma)$ where $\Gamma$ is the ratio of radiation to gravitational forces. With increasing $\Gamma$, the \Lone{} and \Ltwo{} equipotentials come closer to each other until they merge at a critical value $\Gamma_\crit(q)$ and for $\Gamma > \Gamma_\crit$, the star does not possess a contact surface and the \Lone{} equipotential, no longer exists. Similarly to the previous situation with $\mathcal{A} > 1$, the mass leaving \Lone{} can have access to a much larger volume and escape from the system. 
The effect of radiative acceleration from both stars on the Roche lobes was also investigated by \cite{Vanbeveren1977,Salzman1982,Huang1990}, using a different parameter $\Gamma$ for each star. These approaches have limitations however because the shielding effect of the stars (and potentially of the outflow) is not considered. Moreover, if mass transfer takes place in an optically thick regime, the free-stream approximation becomes invalid. As noted previously, under such conditions, the influence of radiation pressure can no longer be modeled using the parameter $\Gamma$ \citep{Howarth1997}. For dust-driven wind in pulsating Mira variables,  the potential becomes periodically repulsive and attractive \citep{Dermine2009}. Formally, this translates into a time dependent parameter $\Gamma$, which on average remains slightly above unity, allowing mass loss.

The effect of irradiation from a companion star has also been investigated \citep{Drechsel1995,Phillips2002,Owocki2007} but the problem is complex due to geometrical effects. Indeed, a careful treatment requires to take into account the angle dependence of the flux hitting the star, producing a non-axisymmetric deformation of the stellar surface and deviation from the standard Roche geometry. Like for asynchronous rotation, the positions of the Lagrangian points are shifted and the existence of an inner contact surface is not warranted. These effects are generally not considered in stellar evolution calculations, because of their complexity but have an impact especially when the system hosts a compact accretor like a neutron star or black hole.

From this discussion, it is clear that the assumptions underlying the Roche model are rather restrictive. Growing observational evidence indicates that the classical Roche formulation requires revision. This is particularly evident in the case of ellipsoidal variables which are short period systems where the shape of the primary star is significantly affected by the gravitational pull of its companion. This interaction distorts the star into an elongated, ellipsoidal shape that nearly fills its Roche lobe. As the system orbits, the viewing angle of the distorted star changes, producing periodic brightness variations known as ellipsoidal light variations.  Recent studies of red giant ellipsoidal variables \citep{Nie2017,Navarrete2025} have revealed that the primary only partially fills its Roche lobe. These findings challenge the predictions of the classical Roche model, unless the size of the Roche lobe is reduced by the effect of additional forces, such as those arising from chromospheric activity or stellar pulsations. This idea was initially proposed by \cite{Frankowski2001} to explain some symbiotic stars and may well apply to these ellipsoidal variables.

An interesting consequence of modifications to the Roche potential is that the \Lone{} equipotential surface may no longer fully enclose the stellar components. This configuration allows for enhanced mass and angular momentum transfer between the stars. In the following section, we address how this effect can be incorporated into binary evolution models.

\subsection{Mass and angular momentum conservation}

The evolution of the orbital parameters is dictated by the conservation of angular momentum and mass. 
The total angular momentum of the binary star system $J_\Sigma$ is the sum of the orbital  ($J_\orb$) and stellar ($J_i$) contributions
\begin{equation}
  J_{\Sigma} = J_\donor + J_\gainer + J_{\orb} \ ,
  \label{eq:Jsigma}
\end{equation}
where
\begin{equation}
    J_\orb = M_\donor M_\gainer\left(\frac{Ga(1-e^{2})}{(M_\donor+M_\gainer)}\right)^{1/2}\ .\label{eq:Jorb}
\end{equation}
Assuming spherical symmetry, the angular momentum of each star writes 
\begin{equation}
J_{i=\donor,\gainer} = \frac{2}{3}\,\int_{0}^{M_{i}}\Omega_{\spin,i}(m)\,r^{2}(m)\,dm\ .
\label{eq:Jstar}
\end{equation} 
Time derivative of Eq.~(\ref{eq:Jorb}) gives the equation for the evolution of the binary separation
\begin{equation}
\frac{\dot{a}}{a}=2\frac{\dot{J}_{\orb}}{J_{\orb}}-2\left(\frac{\dot{M}_\donor}{M_\donor}+\frac{\dot{M}_\gainer}{M_\gainer}\right)+\frac{\dot{M}_\donor+\dot{M}_\gainer}{M_\donor+M_\gainer}+\frac{2e\dot{e}}{1-e^{2}}\ .
\label{eq:adot}
\end{equation}
To evaluate this expression and determine the new separation, prescriptions must be provided for the various mass transfer rates ($\dot{M}_\donor$ and $\dot{M}_\gainer$) and the rates of change of orbital angular momentum ($\dot{J}_{\orb}$) and eccentricity ($\dot{e}$). The value of $\dot{J}_{\orb}$ is given by imposing conservation of the total angular momentum, i.e. from (\ref{eq:Jsigma})
\begin{equation}
\dot{J}_{\orb}=\dot{J}_{\Sigma}-\,\dot{J}_\donor-\,\dot{J}_\gainer \ ,
\label{eq:cons_Jtot}
\end{equation}
where $\,\dot{J}_\donor,\,\dot{J}_\gainer$ are the torques applied to the stars and $\dot{J}_{\Sigma}$ is the rate of angular momentum loss from the system. As will be discussed in the next sections, stellar torques arise from tides, magnetic braking and mass transfer (including both accretion and loss) by wind and RLOF, so that 
\begin{eqnarray}
    \dot{J}_\donor & = & \dot{J}^\tide_\donor + \dot{J}^\mb_\donor + \dot{J}_{\loss,\donor}^\wind + \dot{J}_{\acc,\donor}^\wind+\dot{J}_{\loss,\donor}^\RLOF \ ,\\
    \dot{J}_\gainer & = & \dot{J}^\tide_\gainer + \dot{J}^\mb_\gainer +\dot{J}_{\loss,\gainer}^\wind + \dot{J}_{\acc,\gainer}^\wind+\dot{J}_{\acc,\gainer}^\RLOF \ .
    \label{eq:jdot_star}
\end{eqnarray}

The system's mass, $M=M_\donor+M_\gainer$, varies with the individual stellar mass changes driven by Roche-lobe overflow (Sect.~\ref{sect:RLOF}), wind mass loss, or wind accretion and  writes
\begin{equation}
    \dot{M} = \dot{M}_\donor+\dot{M}_\gainer \le 0\ .
    \label{eq:mdot_sys}
\end{equation}
If the total mass and angular momentum of the system are conserved ($\dot{J}_\Sigma = \dot{M} = 0$), the evolution of the system is said conservative. In this case, $\dot{M}_\donor=-\dot{M}_\gainer<0$ and $\dot{J}_{\orb}=0$, so Eq.~\ref{eq:adot} simplifies to
\begin{equation}
\frac{\dot{a}}{a}=2\frac{\dot{M}_\donor}{M_\donor}(1-q)\ .
\label{eq:adot_conservative}
\end{equation}
This relation indicates that when mass is transferred from the more to the less massive component, i.e. when the mass ratio $q=M_\donor/M_\gainer > 1$, the orbit shrinks and expands when $q<1$. 
Non-conservative evolution arises when the mass lost by one star is not entirely accreted by its companion and can be expressed as
\begin{equation}
   \dot{M}_{3-i} = -\beta \,\dot{M}_i \ ,
    \label{eq:msys}
\end{equation}

Mass loss from the system carries away a fraction of the orbital angular momentum at a rate $\dot{J}_\mass$. Additional contributions from gravitational wave emission ($\dot{J}_\gw$), magnetic braking ($\dot{J}_B$) and interactions with a circumbinary disk ($\dot{J}_\disc$), further redistribute angular momentum. The total rate of change of the system's angular momentum, $\dot{J}_\Sigma$, can therefore be expressed as
\begin{equation}
    \dot{J}_\Sigma = \dot{J}_\gw + \dot{J}_\mass + \dot{J}_B\ + \dot{J}_\disc\  .
\end{equation}

In summary, modeling binary evolution requires prescriptions for the stellar torques (Eq.~\ref{eq:jdot_star}), the rates of stellar mass change, the eccentricity evolution (Eq.~\ref{eq:adot}), and the mechanisms that remove angular momentum from the system (Eq.~\ref{eq:cons_Jtot}). The following sections review existing prescriptions and highlight recent developments in these areas.

\subsection{Mass transfer and torques induced by Roche-lobe overflow}
\label{sect:RLOF}
\subsubsection{Mass transfer rate}

Various prescriptions are available to describe the mass transfer rate by RLOF. The most popular models are those of \cite{Ritter1988} and \citet[][hereafter KR90]{Kolb1990}. To estimate the properties of the gas at the \Lone{} point, they integrate the Bernoulli equation along streamlines, assuming the flow is steady. This model distinguishes two regimes, depending on whether the Roche radius is located in the optically thin or tick region of the star. In the \emph{optically thin} case, the region outside the Roche radius is assumed isothermal (equal to the star's effective temperature) and the mass transfer rate can be expressed as \citep{Ritter1988} 
\begin{equation}
\dot{M}^{\RLOF}_\thin=\dot{M}_{0}\exp\left(\frac{R_\donor-R_{\lone,\donor}}{H_{\phot}}\right),\label{eq:Mdot_ritter}
\end{equation}
where $R_\donor$ is the donor's radius, $H_{\phot}$ the pressure scale height at the photosphere and $\dot{M}_{0} < 0$ the mass transfer rate that the donor star would have if it were precisely filling its Roche lobe.\\
This formulation applies to donor stars with a 'sharp' surface  such as main-sequence stars or white dwarfs with a shallow atmosphere ($\tilde{H}_{\phot}/R_\donor\ll1$). 
In case the Roche lobe is located in the \emph{optically thick} region of the donor star, where the optical depth $\tau>2/3$, the flow is assumed to be adiabatic and the expression for the mass transfer rate becomes
\begin{equation}
\dot{M}^{\RLOF}_\thick=\dot{M}_{0}+2\pi F(q)\frac{R_{\lone,\donor}^{3}}{GM_\donor}\int_{P_{\lone}}^{P_{\phot}}\theta(\Gamma_{1})\left(\frac{\mathcal{R}T}{\mu}\right)^{1/2}dP,\label{eq:Mdot_thick}
\end{equation}
where $P_{\phot}$ and $P_{\lone}$ are the pressures at the photosphere and $\lone$ points respectively, $T$ is the temperature, $\mu$ the mean molecular weight and $\theta$ a function of the adiabatic exponent, $\Gamma_{1}=\left(d\ln P/d\ln\rho\right)_{\mathrm{ad}}$  given by
\begin{equation}
\theta(\Gamma_{1})=\Gamma_{1}^{1/2}\left(\frac{2}{\Gamma_{1}+1}\right)^{\frac{\Gamma_{1}+1}{2(\Gamma_{1}-1)}}\ .
\end{equation}
Improvements to this standard prescription have been proposed \citep{Jackson2017,Marchant2021}, using higher order terms in the expansion of the Roche potential at the \Lone{} point to reconstruct the donor surface. These additional terms become relevant when the star significantly overfills its Roche radius \citep{Pavlovskii2015}. Radiation pressure is not negligible in massive stars and can easily be taken into account in the previous formulation by replacing $\mathcal{R}T/\mu$ in Eq.~\ref{eq:Mdot_thick} by $P/\rho$, where $P$ includes both contributions from gas and radiation. \cite{Marchant2021} showed that this new scheme produces slightly higher mass transfer rates and leads to smaller Roche filling factors, which can potentially prevent (unstable) mass transfer by the outer Lagrangian point. 

Recently, \cite{Cehula2023} proposed a new model in which the Roche potential creates a nozzle effect at the inner Lagrangian point, directing the gas flow between the stars. For a realistic equation of state, the mass transfer rate can be higher or lower by a factor up to two compared to the KR90 prescription depending on the situation.
\cite{Ivanova2024} also devise a new scheme that removes the problem associated with the fact that in the KR formalism, different physical assumptions are used to calculate the mass transfer rate in the optically thin and thick regime. Their unified approach gives a mass transfer rate that is different by a factor of a few compared to those of KR. 
\cite{Cehula2026} revisited the hydrodynamics of mass transfer in close binaries by incorporating the influence of radiation pressure on the flow through the inner Lagrange point. The authors show that when radiation pressure becomes significant, it can substantially boost the mass-transfer rate even before the donor has filled its Roche lobe. For low- and intermediate mass stars on the RGB, the increase is modest, not exceeding 20\% of the standard value.

Hydrodynamical simulations also provide valuable insight into the mass loss process from the donor star.  \cite{Ryu2025} analyzed the flow through both the inner and outer Lagrangian points using high‑resolution three‑dimensional simulations. They showed that the Coriolis force breaks axial symmetry, reduces the stream cross section, and shifts flow features toward the donor’s trailing side. Despite these structural asymmetries, which are not considered in classical models, the resulting mass-transfer rates remain within a factor of a few of standard analytic prescriptions across a wide range of mass ratios. Based on these results, the authors also provide dimensionless scaling factors and revised mass-transfer rate prescriptions for use in binary stellar evolution codes.

\subsubsection{Torque on the donor star}

The material lost from the donor at the $\lone$ point has the specific angular momentum of the stellar surface
\begin{equation}
j_\donor=\Omega_{\spin,\donor}R_\donor^{2}\ ,
\end{equation}
where we can reasonably assume $R_\donor=R_{\lone,\donor}$ (if the radius significantly exceeds the Roche radius other physical processes will take over, such overflow from the other Lagrangian points). The torque applied on the donor resulting from the RLOF mass transfer is
\begin{equation}
\dot{J}_{\loss,\donor}^{\RLOF} = \dot{M}^{\RLOF} \,\Omega_{\spin,\donor} R_\donor^{2}\, < 0 \ .
\end{equation}
where $ \dot{M}^{\RLOF}$ is given by Eq.~\ref{eq:Mdot_ritter} or \ref{eq:Mdot_thick}, depending on the location of the Roche radius with respect to the photosphere.

\subsubsection{Torque on the gainer star\label{subsec:ballistic}}

The material falling into the potential well of the secondary can directly impact the surface of the gainer, distribute into an accretion disk, or leave the system. \cite{Ulrich1976} showed that the ballistic trajectories of a particle ejected from the $\lone$ point will reach, at its closest approach, a radius
\begin{equation}
R_{\mathrm{min}} = 0.0425a\left(q\left[1+q\right]\right)^{1/4}\ ,
\end{equation}
where $q=M_\donor/M_\gainer$.

If $R_{\mathrm{min}}>R_\gainer$, the stream goes around the star and after one orbit collides with itself. After multiple collisions, the mass settles into a circular orbit at the circularization radius
\begin{eqnarray}
R_{\circ} & = & \frac{\oorb^{2}l_\gainer^{4}}{GM_\gainer}=a\left(\frac{l_\gainer}{a}\right)^{4}\left(1+q\right)\label{eq:Rcirc}\ ,\\
 & \approx & a\left(0.5-0.227\log q\right)^{4}(1+q)\ ,\nonumber
\end{eqnarray}
where $l_\gainer$ is the distance between the accretor and the Lagrange point $\mathcal{L}_{1}$. Matter initially accumulates in a narrow ring at $r=R_{\circ}$ and later spreads, moving both inward and outward, and an accretion disk forms. At the gainer's surface, the rate of angular momentum deposition, i.e. the applied torque, is simply
\begin{equation}
\dot{J}_{\acc,\gainer}^{\RLOF} = \dot{M}^{\RLOF}_\acc \,\Omega_{\crit,\gainer} R_\gainer^{2} = \dot{M}^{\RLOF}_\acc\, (GM_\gainer R_\gainer)^{1/2}  > 0 \ ,
\label{eq:torque_RLOF_acc_disc}
\end{equation}
 where
\begin{equation}
\Omega_{\crit,\gainer}=(GM_\gainer/R_\gainer^{3})^{1/2}\label{eq:Omega_crit} \ ,
\end{equation}
is the Keplerian angular velocity at the gainer's surface and
\begin{equation}
\dot{M}^{\RLOF}_{\acc} = -\beta_\RLOF \dot{M}^{\RLOF} > 0 \ ,
\label{eq:beta_RLOF}
\end{equation}
the effective mass accretion rate. Here, $\beta_\RLOF <1$ reflects non-conservative mass transfer, which may occur, for example, when the equipotential surfaces open (Sect.~\ref{sect:roche}) or when the accretion rate exceeds the Eddington limit of the gainer. These processes and their consequences for orbital evolution will be further discussed in Sect.~\ref{sect:eccentric}.

\begin{figure}
    \centering
    \includegraphics[width=\columnwidth]{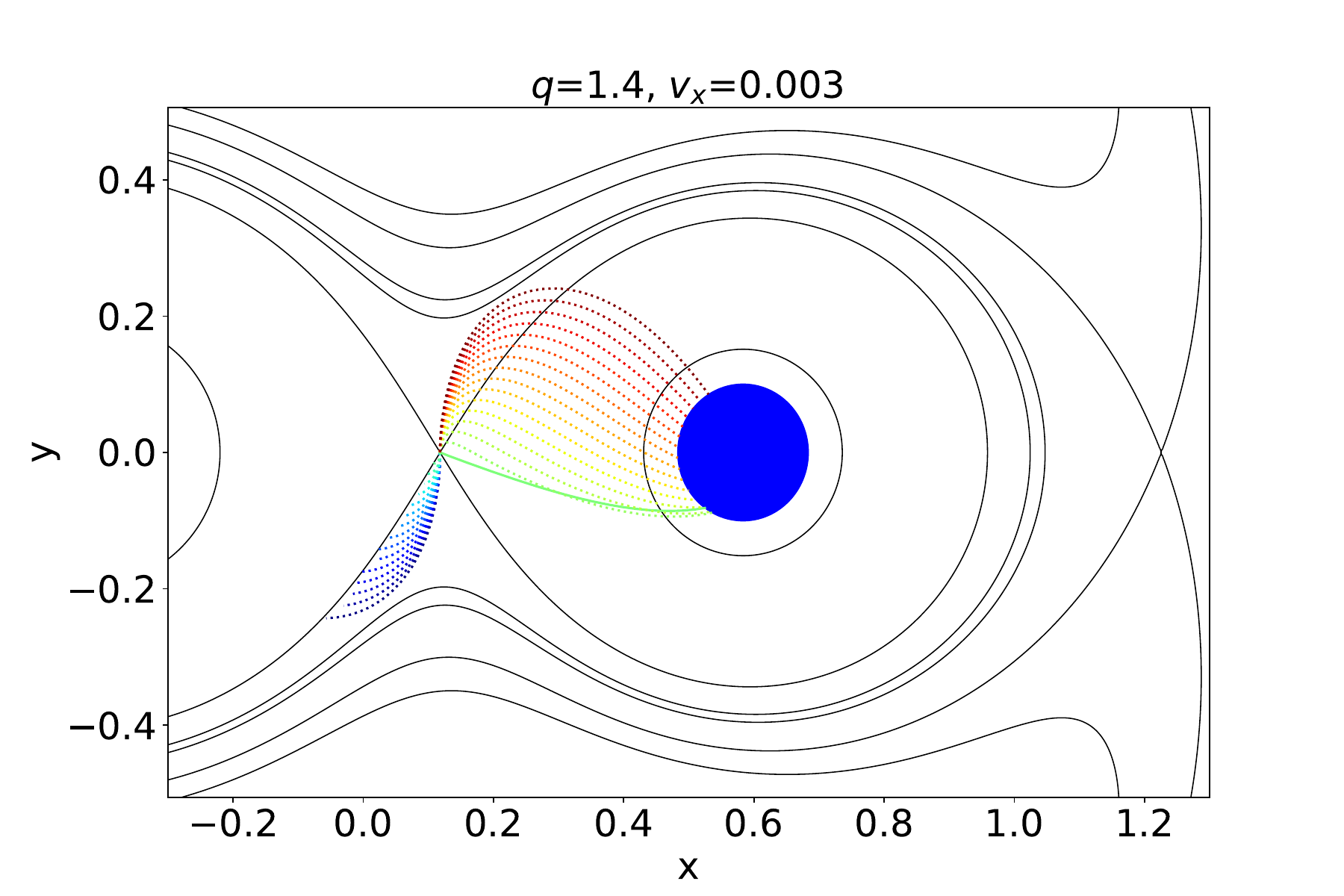}
    \caption{Ballistic trajectories of material escaping the \Lone{} point for different tangential velocity components. The system mass ratio $q=1.4$ and the radial velocity at the \Lone{} point is $0.003\times v_\orb$. The different curves show the trajectories for various tangential velocities, with $f_\rot$ varying from -0.8 to 0.8 in step of 0.1  from blue to red. The thick solid green line corresponds to the synchronous case ($f_\rot = v_y=0$). Coordinates are expressed in the center of mass reference frame. }
    \label{fig:ballistic}
\end{figure}

Alternatively, if $R_\gainer>R_\mathrm{min}$ the stream emanating from the \Lone{} point directly impacts the accretor. Estimating the torque in this case requires knowledge of the direction and magnitude of the stream velocity, which can be obtained by solving the ballistic trajectory of a test particle within the gainer's Roche potential. In the barycentric reference frame corotating with the binary \citep{Flannery1975}, the equation of motion ($\vec{a} = -\nabla \phi_\Roche$) writes
\begin{eqnarray}
\ddot{x} & = & 2\dot{y}+x-\mu\frac{(x-x_\gainer)}{r_\gainer^{3}}-(1-\mu)\frac{(x-x_\donor)}{r_\donor^{3}} \ ,
\label{eq:traj_x} \\
\ddot{y} & = & -2\dot{x}+y-\mu\frac{y}{r_\gainer^{3}}-(1-\mu)\frac{y}{r_\donor^{3}}\ ,
\label{eq:traj_y}
\end{eqnarray}
where the test particle is located at $\vec{r}=(x,y)$, the donor at $\ensuremath{x_\donor=\mu-1}$, the gainer at $x_\gainer=\mu$. Here $r_\gainer$ and $r_\donor$ denote the particle's distances from the accretor and donor, respectively (see Fig.~\ref{fig:notation}). In the above equations, the first term proportional to the particle's velocity corresponds to the Coriolis forces. The equations are formulated in a non-dimensional form in which distances are expressed in units of orbital separation ($a$), time in units of orbital period ($2\pi/\oorb$) and masses in units of reduced mass $\mu=\frac{M_\gainer}{(M_\donor+M_\gainer)}$.
When the particles collide with the accretor, its specific angular momentum is given by 
\begin{eqnarray}
j_{\impact} & = & \left\Vert \,\vec{r}\wedge\left(\vec{v}+\vec{\Omega}_{\orb}\wedge\vec{r}\right)\,\right\Vert \label{eq:jacc} \ ,\\
 & = & \left[(x-x_\gainer)\dot{y}-y\dot{x}+r_\gainer^{2}\right]a^{2}\oorb\ ,
\end{eqnarray}
from which we obtain the torque
\begin{equation}
\dot{J}_{\acc,\gainer}^{\RLOF}=  \dot{M}^{\RLOF}_\acc \,j_{\impact}\, > 0\ .
\label{eq:torque_RLOF_acc_impact}
\end{equation}
The ballistic trajectories are sensitive to the initial velocity of the stream. Notably, in systems where the stars are not rotating synchronously, a tangential velocity component naturally arises. Concerning the radial component, it is of the order of the thermal velocity of the photospheric layers and thus depends on the physical characteristics of the donor star. A systematic study of the mass stream was investigated by \cite{Hendriks2023}. These authors showed that if the overfilling star rotates asynchronously, a fraction of the ejected particles can fall back onto the star \citep{Kruszewski1964} and the amount of re-accretion increases with decreasing initial stream velocity. Figure \ref{fig:ballistic} illustrates the particle trajectories, launched from \Lone{} with a radial velocity $v_x  = 0.003\times v_\orb$ and a tangential component $v_y = f_\rot \times v_\orb$ with $f_\rot$ varying between -0.8 to 0.8. For sub-synchronous rotation ($f_\rot <0$), the particles fall back onto the donor star, inhibiting accretion on the companion. 

When the stream impacts the stellar surface or the outer edge of an accretion disk, a hotspot forms where the stream's kinetic energy is converted into heat, locally increasing the luminosity. If the hotspot luminosity exceeds the Eddington limit, matter can be expelled from the system. This mechanism has been investigated by \cite{Rensbergen2008,Rensbergen2010,Rensbergen2011} and \cite{Deschamps2013} in the context of short-period Algol systems, in which mass transfer occurs either during the donor's main sequence (case A) or as it crosses the Hertzsprung gap (case B). These studies show that mass transfer remains conservative below a critical accretion rate. However, once this threshold is exceeded, up to 50\% of the transferred material can be ejected. Mass ejection predominantly occurs during the rapid phase of mass transfer and depends on the initial separation and the adopted opacity at the surface of the gainer star where the Eddington luminosity is evaluated. By driving non-conservative evolution, this process helps reproduce the observed mass ratio distribution of Algol systems. 

\subsection{Wind mass loss}
\label{sect:wind_loss}

Because of the strong coupling between the (ionized) stellar plasma and magnetic field, the angular momentum carried away by a stellar wind is highly sensitive to magnetic activity. We therefore first examine the non-magnetic case, before turning to magnetized winds in the following sections.

\subsubsection{Non-magnetic stars}

A mass element expelled in the wind of a rotating star removes the specific angular momentum of the stellar surface layers. For isotropic ejection, the wind can be modeled as the release of a spherical shell, and the rate at which angular momentum is lost writes 
\begin{equation}
\dot{J}_{\loss,i}^{\wind}=\,\frac{2}{3} \dot{M}_{i}^{\wind}\,\Omega_{\spin,i}\,R_{i}^{2}\, < 0\ ,
\label{eq:torque_wind_loss}
\end{equation}
where $\dot{M}_{i}^{\wind}$ is the wind mass loss rate from star $i$ which depends on the stellar parameters (mass, effective temperature, metallicity, ..) and evolutionary status. In this expression, the orbital angular momentum carried away by the wind is not included, but its contribution is accounted for separately through $\dot{J}_\mass$.

Stellar rotation also influences wind mass loss. In particular, if the star approaches critical velocities, as a result of contraction or accretion of angular momentum for example, the mass loss rate can be significantly enhanced ($\Omega$ limit) and corrections should be applied. The same holds for very luminous stars that approach the Eddington limit ($\Gamma$ limit). These aspects have been discussed in detail in \cite{Maeder2000}. 

Rotation induces a deformation of the stellar structure through centrifugal forces, and because the radiative flux is proportional to the effective gravity (Von Zeipel theorem), it varies across the stellar surface. As a consequence, radiatively driven winds are further accelerated in the polar direction creating equatorial density enhancement and anisotropic mass ejection. Recent work on massive stars \citep{Hastings2023} highlights the complexity of these effects, but also  reveals that rotation does not necessarily lead to stronger mass loss.

In binary systems, the presence of a companion can further alter wind mass loss. \cite{ToutEggleton1988}  suggested that tidal interactions and dynamo activity could enhance stellar winds, and proposed a prescription where the mass-loss rate scales with stellar radius and orbital separation in the same way as the tidal torque acting on the star, i.e. 
\begin{equation}
    \dot{M}_i^\wind = \dot{M}^\wind_{0,i}\left[1+\alpha^\wind_i\times\min\left(\frac{R_i}{R_{\lone,i}},\frac{1}{2}\right)^6 \right] \ ,
    \label{eq:CRAP}
\end{equation}
where $\alpha^\wind_i$ is a free parameter and $\dot{M}^\wind_{0,i}$ the standard mass loss rate. In this formulation, the tidal interaction increases until $R = \frac{1}{2}R_\lone$ where synchronization is assumed to be reached. This simple model was used to interpret a variety of binary systems, including Algols \citep{ToutEggleton1988,Zhang2013}, type Ia supernovae \citep{Chen2011}, eccentric binaries hosting a helium white dwarf \citep{Siess2014}, CH and Ba stars \citep{Han1995b,Escorza2020,Krynski2025},  the horizontal branch morphology of globular clusters \citep{Lei2013} or hot sub-dwarfs \citep{Vos2015}. Despite successes, this prescription lacks a solid physical foundation. It should also be stressed that in an eccentric orbit the mass loss rate becomes phase dependent because the Roche radius \RL{} depends on the instantaneous
separation  (Eq.~\ref{eq:inst_sep}). As we will see in Sect.~\ref{sect:eccentric}, this could further affect the orbital elements.

\subsubsection{Stars with magnetic fields}
\label{sect:magnetic}

In a non-magnetic stellar wind, angular momentum is carried away from the star with a lever arm comparable to the stellar radius (Eq.~\ref{eq:torque_wind_loss}). The presence of a magnetic field, however, enforces co-rotation of the outflow with the stellar surface out to the Alfv\'en radius ($R_A$), where the magnetic and wind kinetic energy density balance each other. Beyond $R_A$, the wind decouples from the magnetic field and flows radially, but each particle carries away the angular momentum as if launched from $R_A$ rather than from $R_i$ \citep{Weber1967}. The corresponding torque is
\begin{equation}
    \dot{J}_{B,i}^{\wind}=\,\frac{2}{3} \dot{M}_{i}^{\wind}\,\Omega_{\spin,i}\,R_{A}^{2}\, < 0\ .
\label{eq:torque_wind_loss_mag}
\end{equation}
where the Alfv\'en radius can be expressed as
\begin{equation}
    R_{A}=\left[(-\dot{M}_{i}^{\wind})^{-2}B_{i}^{4}(2GM_i)^{-1}R_i^{4n}\right]^{1/(4n-5)}\ ,
    \label{eq:Alfven}
\end{equation}
where $B_i$ is the magnetic field at the stellar surface and $n$ a parameter that describes the topology of the magnetic field, the case $n=3$ corresponding to a dipole field \citep{Dervisoglu2010,Deschamps2013}. The torque is very efficient at braking the star because usually $R_A \gg R_i$. 

When a star is surrounded by an accretion disk, its magnetic field can anchor in the disk. Differential rotation between the star and the disk twists the field lines, generating a toroidal component that exerts a torque on the star. The sign of the torque depends on whether the magnetic field is anchored inside or beyond the corotation radius leading to either a spin up or spin down of the star. \cite{Stepien2000} showed that the breaking is maximum when all the field lines anchor beyond the corotation radius. In these conditions, the torque is independent of the disk properties (e.g. its mass) and is given by  
\begin{equation}
\dot{J}_{B,i}^{\disc}=-\frac{\mu_B^2\Omega_{\spin,i}^{2}}{3GM_i} < 0\ .
\end{equation}
where $\mu_B= B_iR_i^3$ is the magnetic moment.

Magnetic braking is an efficient mechanism for removing angular momentum, particularly when the donor star possesses a convective envelope capable of sustaining dynamo activity \citep{Tout1992}. It plays a key role in spinning down the accretor in binaries that have undergone Roche-lobe overflow. As demonstrated by \cite{Packet1981}, accretion of only a few percent of its own mass is sufficient to spin the gainer up to near-breakup velocity. This theoretical expectation, however, contrasts with observations of Algol systems, where gainers are typically found close to synchronous rotation despite having accreted substantial amounts of mass. 
Evolutionary models that account for both non-conservative mass transfer and magnetic braking provide a more accurate representation of the observed distribution of equatorial velocities among Algol gainers, aligning theoretical predictions with empirical data \citep{Rensbergen2020,Rensbergen2021}.

Magnetic fields also play a crucial role in the evolution of cataclysmic variables, which are close binary systems where a white dwarf accretes material from a low-mass donor star via Roche-lobe overflow. In these systems, the orbital separation is small enough that strong tidal torques  enforce synchronous rotation of the donor star. This tidal locking couples the stellar spin to the orbital motion, so that angular‑momentum loss from the star is extracted directly from the orbit. Because the orbital angular velocity in close binaries is much higher than the spin rate of an isolated star of comparable mass, magnetic braking (which depends on the stellar spin rate) acts as an efficient sink of orbital angular momentum  \citep{Verbunt1981}. Magnetic braking therefore drives angular momentum loss and thereby regulates the mass transfer rate. This mechanism also accounts for the observed ``period gap'', the dearth of CVs with orbital periods between approximately 2 and 3 hours. Within this period range, mass transfer ceases as the donor becomes fully convective and magnetic braking is disrupted. During this detached phase, angular momentum continues to be removed through gravitational radiation until contact is re-established near  $P_\orb \approx 2$~hr, at which point the system resumes mass transfer and reappears as an active CV. Numerous prescriptions for magnetic braking have been proposed and for a comprehensive overview of the most widely used implementations, we refer the reader to \cite{Knigge2011}.

\subsection{Wind accretion}
\label{sect:wind_acc}

In binary systems, stars travel through the expelled wind of their companion and can accrete mass and angular momentum. In the following sections, we review some idealized cases and recent results from hydrodynamical simulations \cite[for a recent review][]{Boffin2015}. 

\subsubsection{Bondi-Hoyle-Lyttleton model}

When an object of mass $M$ moves at supersonic velocity $v$ through a uniform medium of density $\rho$, it can accrete mass at a rate \citep{Hoyle1939}  
\begin{equation}
    \dot{M}_\mathrm{HL} = \frac{4\pi (GM)^2}{v^3} \rho = \pi R_\acc^2\, \rho v\ ,
    \label{eq:HL}
\end{equation}
where the accretion radius is defined as
\begin{equation}
    R_\acc = \frac{2GM}{v^2}\ .
    \label{eq:Racc}
\end{equation}
This derivation is solely based on the conservation of linear and angular momentum and  pressure terms are neglected. Due to symmetry, the velocity of particles colliding behind the central object aligns with its direction of motion. If this velocity component, determined by the impact parameter, is less than the star's escape velocity, the particle is accreted. This idealized model was later refined by \cite{Bondi1944}, who demonstrated that Eq.~\ref{eq:HL} represents an upper bound.

If the central object is now at rest and material steadily falls onto it, it will accelerate and eventually become supersonic before it reaches the surface. Assuming the gas is at rest at infinity and a polytropic equation of state, \cite{Bondi1952} derived the following expression for the spherical mass accretion rate onto the central object 
\begin{equation}
     \dot{M}_\mathrm{B} = \alpha \pi R_\mathrm{B}^2\, \rho c_s \ ,
     \label{eq:Bondi}
\end{equation}
where $\alpha$ is a parameter that depends on the equation of state (polytropic index) and the Bondi radius $R_\mathrm{B}$ has the same expression as Eq.~\ref{eq:Racc} where the velocity is replaced by the sound speed $c_s$. The transonic solution selected by \cite{Bondi1952} is unique and yields the maximum accretion rate onto the central object \citep{Aguayo2021}. Equations \ref{eq:HL} and \ref{eq:Bondi} are closely related, with the Bondi model applying to stationary stars relative to the ambient gas, while the Hoyle-Lyttleton model describes objects moving supersonically through interstellar material. The Bondi-Hoyle-Lyttleton accretion rate combines these two idealized scenarios \cite[for a broad review on the topic, see][]{Edgar2004} in a single prescription
\begin{equation}
     \dot{M}_\BHL = \alpha \pi R_\acc^2\, \rho v \left(\frac{\Mach^2}{1+\Mach^2}\right)^{3/2} \ ,
     \label{eq:BHL}
\end{equation}
where $\Mach = v/c_s$ is the Mach number.

In binary systems, when accretion occurs through capture of wind material from the companion (referred here as star with index $3-i$), the wind density can be estimated using the mass conservation equation ($\dot{M}^\wind_{3-i} = 4\pi d^2 \rho v^\wind_{3-i}$). Substituting this expression in Eq.~\ref{eq:BHL} and replacing the outflow velocity by the sum of orbital and wind velocities,  
the Bondi-Hoyle-Littleton (BHL) accretion efficiency writes
\begin{equation}
\begin{split}
\beta_{\BHL} = & -\frac{\dot{M}_{\acc,i}^{\wind}}{\dot{M}_{3-i}^{\wind}} \ge 0 \ , \\
= &\ \alpha_\BHL \left(\frac{a}{d}\right)^2 \left(\frac{M_i}{M_1+M_2}\right)^2
\frac{ \left(\displaystyle\frac{v_\orb}{v^\wind_{3-i}}\right)^4}{\left(\displaystyle1+ \frac{v_\orb^2}{{v^\wind_{3-i}}^2} + \frac{c_s^2}{{v^\wind_{3-i}}^2}\right)^{3/2}}\ ,
\label{eq:beta_BH}
\end{split}
\end{equation}
where $\alpha_{\BHL}$ is a free parameter of order unity, $v_\orb$ the orbital velocity 
\begin{equation}
v_{\orb}^2 = \frac{G(M_{\donor}+M_{\gainer})}{a}\ ,
\end{equation}
and $v^\wind_{3-i}$ the wind velocity that can be conveniently prescribed in terms of the escape velocity \citep{Hurley2002} 
\begin{equation}
v^\wind_{3-i} =  \sqrt{2\beta_{w}\frac{GM_{3-i}}{R_{3-i}}}\ ,
\end{equation}
with $\beta_{w} \sim 1/8$ for giant stars with slow winds. In eccentric systems, the BHL accretion efficiency varies with the orbital phase because of its dependence on the instantaneous separation $d$ (Eq.~\ref{eq:inst_sep}). Assuming constant wind velocity and mass loss rate at the accretor's location, the phase-average value of $\beta_\BHL$ can be obtained by replacing the term $(a/d)^2$ by $1/\sqrt{1-e^2}$ \citep{Boffin1988}.

It is important to emphasize that Eq.~\ref{eq:beta_BH} is valid only for \emph{fast winds} ($v_{\wind}\gg v_{\orb}$) and provides a crude estimate of the accretion efficiency if the orbit is eccentric. 
From an analytical standpoint, in the limit $v_\orb \gg v_\wind, c_s$, Eq.~\ref{eq:beta_BH} predicts $\beta_{\BHL} \propto v_\orb/v_\wind$, which can exceed unity, an unphysical result.
More recently, \cite{Tejeda2025} derived a new prescription for the BHL accretion efficiency, introducing a geometric factor that accounts for the orientation of the wind direction with respect to the orbital motion of the star. Their formulation is better in agreement with hydrodynamical simulations and removes the problem of $\beta_\BHL >1$.  \cite{Maldonado2025} used this formalism to study wind accretion in symbiotic systems where a white dwarf (mass between 0.7 and 1\msun) accretes material from the wind of a distant low- or intermediate-mass giant companion. Their study shows that with the new BHL formalism none of their white dwarfs could reach the Chandrasekhar mass, contrary to models using the standard BHL prescription. The treatment of wind accretion can thus have a strong impact on the evolution of symbiotic systems and on the formation of SNIa via this channel. 

\citet{Morris1990} and \citet{Han1995} introduced the gravitational focusing parameter $\alpha_\foc$, equivalent to $\beta_\BHL$ when neglecting the contributions of sound-speed and eccentricity and setting $\alpha_\BHL=1$. They find that small values ($\alpha_\foc \ll 0.1$) correspond to nearly spherically symmetric outflows, while larger values ($\alpha_\foc \gtrsim 0.1$) indicate strong focusing and the emergence of bipolar structures. This link between accretion efficiency and wind geometry highlights the need for hydrodynamical treatments that go beyond the simplistic fast-wind assumption of the BHL model.

\subsubsection{Results from hydrodynamical simulations}

\begin{figure*}
    \centering
    \includegraphics[width=\linewidth]{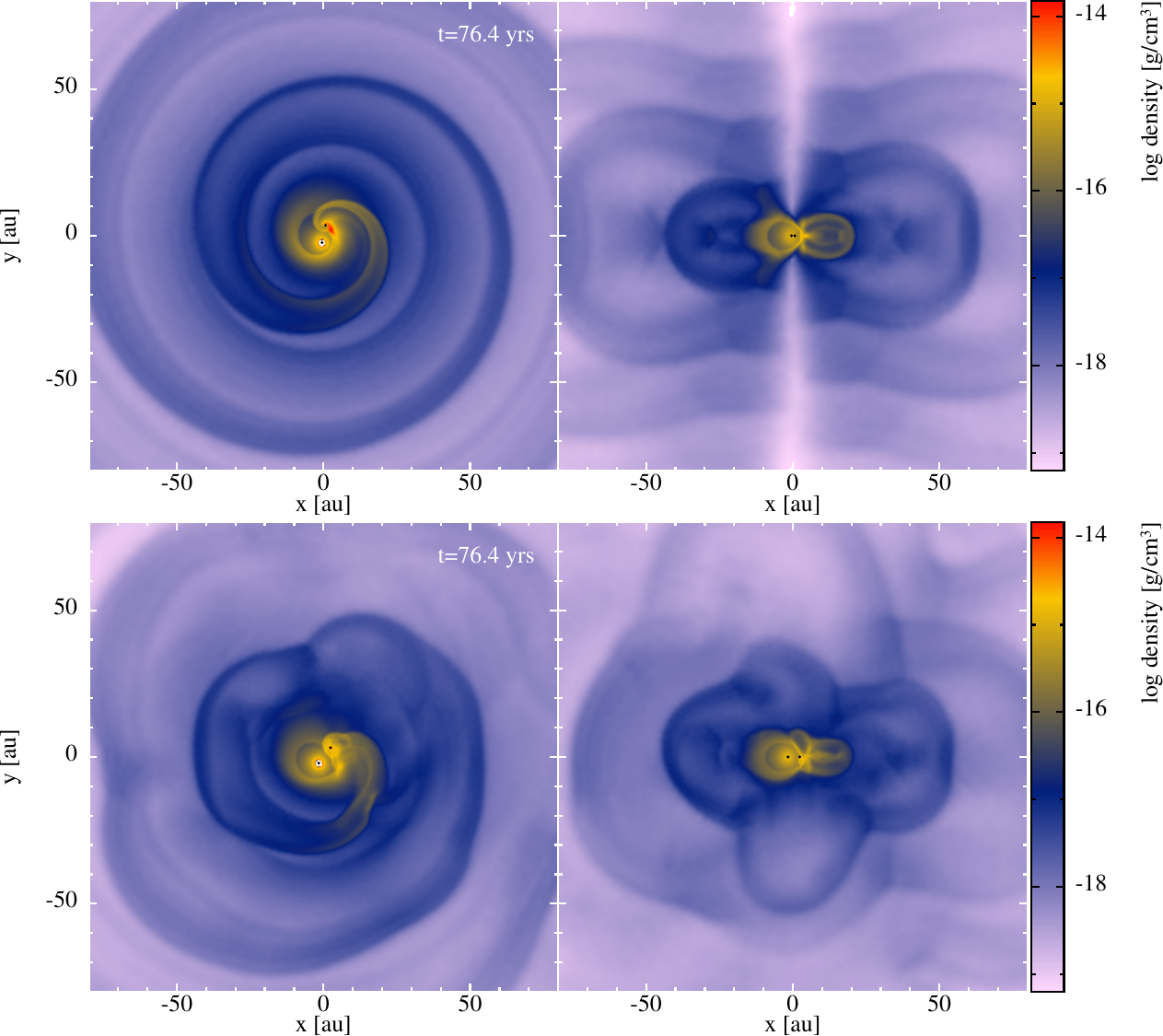}
    \caption{Two dimensional slice revealing the density structure in the equatorial (left column) and meridional (right column) plane for binary wind interaction in a circular (top row) and eccentric (bottom row) orbit. The AGB star has a mass of 1.5\msun, an effective temperature of 3000~K and loses mass at a rate of $10^{-7}$\myr. The companion has a mass of 1.2\msun, the orbital separation is 6~au and in the bottom row the eccentricity $e=0.25$. Note the strong dependence of the patterns on the viewing angle.}
    \label{fig:sph}
\end{figure*}

The presence of a binary companion can have a profound impact on the morphology of the ejected wind material \citep{Theuns1993,Mastrodemos1998,deValBorro2009,Liu2017,Chen2017,Saladino2019,Malfait2024}. When a companion star orbits within the stellar wind of a mass-losing primary, such as an AGB star, its gravity disturbs the outflowing gas, creating a dense wake. As the companion moves along its orbit, this wake follows its motion and gradually wraps around the system, forming a spiral pattern. The shape and spacing of the spiral depend on the wind speed, the orbital velocity, and the distance between the stars. Slower winds and compact systems produce tighter spirals, while faster winds lead to more widely spaced spirals. The presence of a companion also focuses the wind toward the orbital plane \citep[e.g.][]{Maes2021}. For slow winds, the gravitational pull is stronger, producing larger equatorial densities and a wider spiral arm separation. At high velocities, the flow approaches the BHL regime and the spiral arms are much tighter because of the smaller accretion radius. The flow also remains more spherical in this case. Figure \ref{fig:sph} illustrates the wind morphology in a binary system consisting of a 1.5\msun{} AGB donor and a 1.2\msun{} companion. The top panel shows the structure produced in a circular orbit, while the bottom panel corresponds to an eccentric configuration with $e = 0.3$. In both cases, the systems share the same semi-major axis of $a = 6$ au. As the companion moves through the AGB wind, it generates a bow shock that deflects the outflow, producing a regular spiral pattern in the circular case. In contrast, orbital eccentricity induces an asymmetric and more intricate morphology \citep[for details, see][]{Malfait2021}. The patterns emerging from the hydrodynamical simulations are found to depend sensitively on the viewing angle, exhibiting a wide variety of appearances such as spirals, arcs, or shell-like structures whose degree of elongation and regularity varies across different orientations. Comparable features have been identified in high-resolution observations of AGB stars \citep[e.g.][]{Maercker2012,Randall2020,Decin2020}, providing indirect evidence of the presence of a companion.

Asymmetries in stellar outflows can strongly affect the mass accretion rate onto the companion star, often leading to significant departures from the classical Bondi-Hoyle-Lyttleton prescription. These discrepancies are especially marked in systems with slow stellar winds and in environments where dust formation occurs within the outflow. To understand the origin of such deviations, it is essential to examine the physical processes that govern mass loss in evolved stars and the conditions under which these winds interact with a binary companion. In binaries involving an Asymptotic Giant Branch (AGB), a distinct mode of mass transfer mode, known as Wind Roche-Lobe Overflow (WRLOF), can arise \citep{Podsiadlowski2007,MohamedPodsiadlowski2007,Mohamed2010}. Mass loss in AGB stars is a complex phenomenon. Stellar pulsations and large-scale convective motions generate shock waves that propagate through the atmosphere, lifting material to cooler regions where dust can  condensate. Radiation pressure acting on the newly formed dust grains then accelerates the flow, with the gas entrained through collisions with the grains \citep{Hofner2018}. 
The dust condensation temperature depends primarily on the chemical composition of the outflow, particularly the carbon-to-oxygen (C/O) ratio. For oxygen-rich (O-rich) stars, dust typically condenses at temperatures of $\sim 1000$~K, while in carbon-rich (C-rich) stars, condensation occurs at higher temperatures, up to $\sim 1500$~K. Consequently, the dust formation radius, $R_\cond$, is sensitive to both the composition and the density structure of the circumstellar envelope.
In binary systems, the relative location of $R_\cond$ with respect to the Roche lobe radius \RL{} plays a critical role in shaping wind interaction. If \RL{} lies near or inside $R_\cond$, the wind velocity remains low within the Roche lobe. Consequently, the companion's gravitational influence is  more pronounced, and matter can efficiently channel toward the secondary, hence the term wind Roche lobe overflow. Conversely, if $R_\cond$ is significantly smaller than \RL, the wind material will acquire a higher velocity when it reaches the Roche lobe and the interaction with the companion will be weaker. In this case, the accretion rate onto the secondary is expected to be closer to the BHL values. 

\begin{figure}
    \centering
    \includegraphics[width=\linewidth]{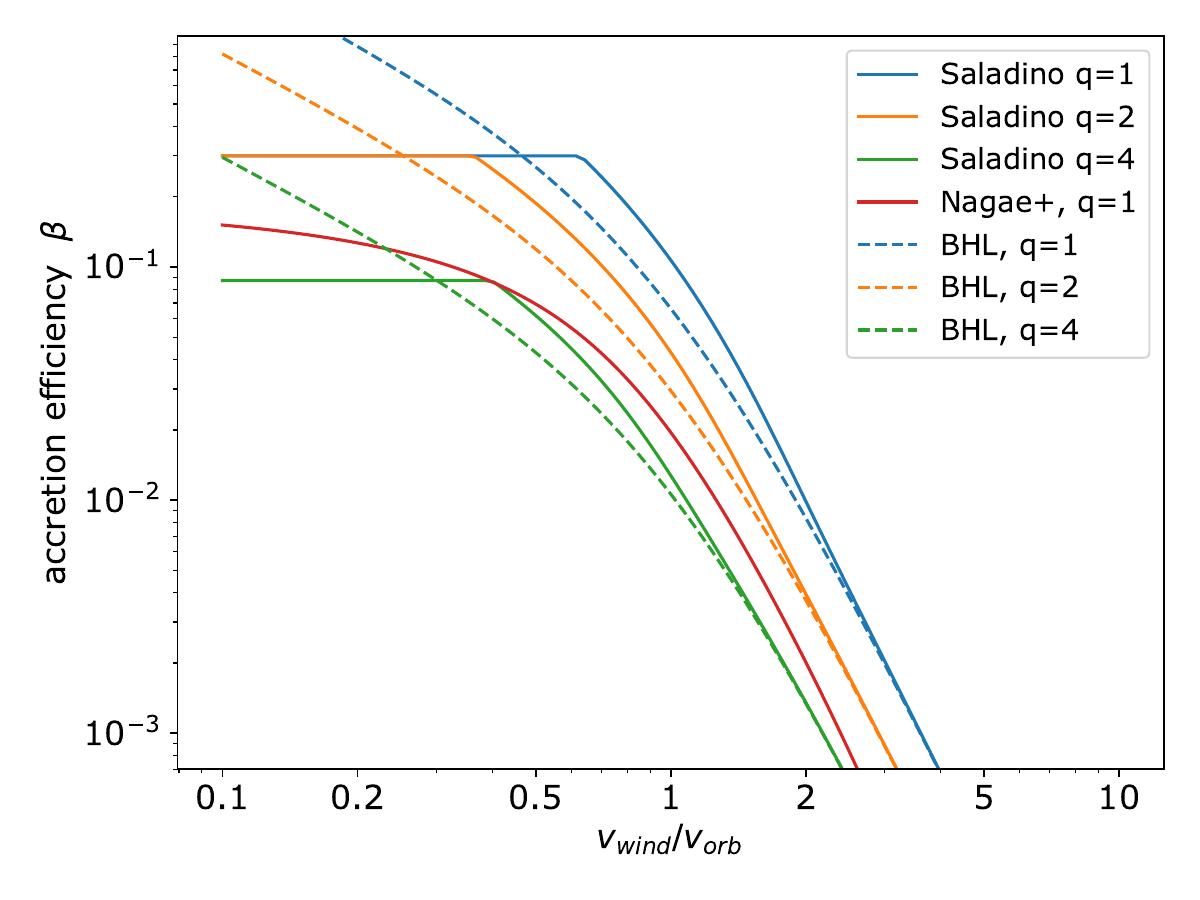}
    \caption{Accretion efficiency $\beta$  as a function of the  wind-to-orbital velocity ratio for different available prescriptions. Solid lines represent values computed using the empirical relation from \cite{Saladino2019} (Eq.~\ref{eq:beta_saladino}) for different mass ratios $q$. The dashed lines show the corresponding predictions from the BHL formalism (Eq.~\ref{eq:beta_BH}). The red solid line shows the fit derived by \cite{Nagae2004} for $q=1$.}
    \label{fig:beta_acc}
\end{figure}

\cite{Saladino2019ecc} conducted an extensive series of smoothed particle hydrodynamics (SPH) simulations to estimate the accretion efficiency in a binary system hosting a mass-losing AGB primary. They varied the mass ratio, orbital separation, initial wind velocity and donor star's rotation rate. Their models considered an evolved AGB star of 1.2\msun{} and 330\rsun, with a mass loss rate of $1.5\times 10^{-5}$\myr and included a simple prescription for dust opacity based on \cite{Bowen1988}. They derived the following fit to the accretion efficiency 
\begin{equation}
\beta_\wind = -\frac{\dot{M}^\wind_{\acc,i}}{\dot{M}^\wind_{3-i}} =  \min(\alpha \beta_\BHL,\beta_{\max})\ ,
\label{eq:beta_saladino}
\end{equation}
where
\begin{equation}
\alpha = 0.75+\frac{1}{k_1+(k_2v_\wind/v_\orb)^5}\ ,
\end{equation}
with $k_1 = 1.7+0.3\,q$, $k_2=0.5+0.2\,q$, $\beta_{\max} = \min(0.3,1.4\,q^{-2})$ and $\beta_\BHL$ given by Eq.~\ref{eq:beta_BH} with $\alpha_\BHL =1$ and $q = M_\donor/M_\gainer$. Figure \ref{fig:beta_acc} shows the fitted relation derived from their simulations together with the values predicted by the BHL formalism for a range of mass ratios. The results indicate that the mass–accretion rate increases with decreasing $v_\wind/v_\orb$ ratio  and that for $v_\wind/v_\orb > 0.5$, the accretion efficiency in their models consistently exceeds the BHL prediction ($\beta_\wind > \beta_\BHL$).
Comparison of their results with \cite{Mohamed2010}, \cite{Abate2013} and \cite{Chen2017} reveals a consistent trend of decreasing accretion efficiency with increasing wind-to-orbital velocity ratio ($v_\wind/v_\orb$). Their simulations also align with those of \cite{Theuns1996} and \cite{Liu2017}, showing that the inclusion of gas cooling substantially enhances the accretion efficiency. However, notable discrepancies exist among the models (see their Fig. 9). For instance, \cite{MohamedPodsiadlowski2007} report high accretion efficiencies under the WRLOF regime, in contrast to \cite{Saladino2019} whose results show modest enhancements compared to the classical BHL predictions in similar conditions.
These discrepancies are attributed to differences in input physics (cooling, dust, spin of the donor star, orbital parameters), numerical resolution, and to the algorithm used to estimate the accretion rate but the main features are captured. 

Observationally, high wind–accretion efficiencies are required to reproduce the surface abundance patterns observed in chemically peculiar binary systems such as CEMP-s, CH, and barium stars \citep[e.g.][]{Bisterzo2012,Cseh2022}. These systems exhibit pronounced overabundances of heavy nuclei (in particular s-process elements like barium), which are generally attributed to mass transfer from a former AGB companion via its stellar wind. A scenario involving Roche-lobe overflow  from an AGB donor is considered unlikely, as it would typically result in common-envelope evolution, leading either to a merger or the formation of a short-period binary. In a recent study, \cite{Dimoff2025} suggest that between 0.1 and 0.5~\msun{} must be accreted to explain the observed enrichment. Such a large amount of accreted material remains difficult to reconcile with wind accretion alone, unless RLOF could proceed in a stable manner.

\subsubsection{Stellar torque due to wind accretion}

When the stellar wind impacts the companion star, it also transfers angular momentum. However, due to the complexity of the flow dynamics, quantifying this transfer is challenging. In the idealized Bondi-Hoyle accretion scenario, the net angular momentum input may vanish \citep{Ruffert1994a, Ruffert1994b, Blondin2012}. Yet, in systems with eccentric orbits or anisotropic wind emission, the outcome can differ significantly. \citet{Shapiro1976} and \citet{Jeffries-Stevens-96} parameterized the torque exerted by the wind of the primary onto its companion as
\begin{equation}
\dot{J}_{\acc,i}^{\wind}=\frac{1}{2}f_{\mathrm{Jacc}}\dot{M}^{\wind}_{\acc,i}\,\oorb R_{\acc,i}^{2} > 0\ ,
\label{eq:torque_wind_acc}
\end{equation}
where $f_{\mathrm{Jacc}}$ is a free parameter that varies between $0.01-0.1$ and $R_{\acc,i}$ the accretion radius of the wind accreting star 
\begin{equation}
R_{\acc,i}=\frac{2GM_i}{v_{\orb}^{2}+\left(v_{\wind}^{2}\right)} \ .
\end{equation}
If the specific angular momentum of the accreted material, $j_\acc = \dot{J}_{\acc,i}/\dot{M}_{\wind,i}^{\acc}$, exceeds the Keplerian value at the surface of the gainer, an accretion disk is expected to form around the companion star. In this case, the disk is supplied not only by the direct gas stream from the donor but also by material channeled through the leading spiral arm structure ahead of the star. Disk formation depends sensitively on the wind velocity relative to the orbital velocity as well as on the adopted equation of state. SPH simulations by \cite{Theuns1996} show that adopting an adiabatic equation of state suppresses disk formation because the high temperature in the vicinity of the companion produces an overpressure that prevents direct accretion. In contrast, the inclusion of cooling processes \citep{Malfait2024} favors disk formation. In the absence of more realistic prescriptions, binary evolution models have to resort on the use of the parametric Eq.~\ref{eq:torque_wind_acc}.

\subsection{Systemic angular momentum loss}
\label{sect:systemic}

When mass is ejected from a star and leaves the system, it carries away some of the system's orbital angular momentum. A precise estimate of this quantity requires hydrodynamical simulations that can capture the complex interactions between the outflowing material and the binary components. However, in some idealized scenarios, analytical expressions for angular momentum loss can be derived \citep{Huang1956,Huang1963,Soberman1997}. 

If the mass is ejected isotropically and does not interact with the companion star, it takes away the orbital angular momentum at the star's location on the orbit. This mode of mass ejection is referred to as rapid mode or Jeans mode and in this case, the rate at which angular momentum is removed from the system by the mass losing star writes
\begin{equation}
\dot{J}^\jeans_{i} =\dot{M}_{i}^{\wind} \frac{M_{3-i}}{M_i}\frac{J_\orb}{M} = \dot{M}^\wind_i a_i^2 \oorb < 0\ ,
\label{eq:Jdot_Jeans}
\end{equation}
where $a_i$ is the distance of star $i$ to the center of mass. 
If both stars are in this wind mass loss regime, the systemic angular momentum loss rate is 
\begin{eqnarray}
  \dot{J}_{\mass}^\jeans & = & \dot{M}_\donor^{\wind}a_\donor^{2}\oorb + \dot{M}_\gainer^{\wind}a_\gainer^{2}\oorb \nonumber \\
& = &   \left[\frac{\dot{M}_\donor^{\wind}}{q}+\dot{M}_\gainer^{\wind}q\right]\frac{M_\donor M_\gainer}{M^{2}}a^{2}\oorb \ ,
\label{eq:Jdot_sys_Jeans}
\end{eqnarray}
where the mass ratio $q=M_\donor/M_\gainer$. 

In some circumstances, for example when mass is transferred onto a compact object, the mass accretion rate on the companion can exceed the Eddington limit. In this case, the matter cannot be fully accreted and a fraction is expelled from the vicinity of the accretor. This so-called re-emission mode is relevant when the system hosts an accreting neutron star where the Eddington rate $\dot{M}_\Edd$ is relatively low, of the order of a few $10^{-8}$\myr. In this regime, the same formula as Eq.~\ref{eq:Jdot_Jeans} applies except that mass is now lost by the accretor. This yields the following expression for the rate of systemic angular momentum loss from star $i$
 \begin{equation}
\dot{J}^\rem_\mass =\dot{M}_{i}^{\rem} \frac{M_{i}}{M_{3-i}}\frac{J_\orb}{M} = \dot{M}_{i}^{\rem} a^2_{3-i}\Omega_\orb \ ,
\label{eq:Jdot_remission}
\end{equation}
where $\dot{M}_{i}^{\rem} \le 0$ represents the mass loss rate due to re-emission, whose expression is parameterized by
\begin{equation}
\dot{M}^\rem_i = -\beta^\rem_i\, \dot{M}^\RLOF_{\acc,i} < 0\ .
\end{equation}
The mass accretion rate on the gainer star being limited to the Eddington value,
\begin{equation}
  \dot{M}^\RLOF_{\acc,i} = \min(\dot{M}^\RLOF_{\acc,i},\dot{M}_\Edd)\ .
\end{equation}
we find that  
\begin{equation}
\beta^\rem_i = \min\left(1,\frac{\dot{M}_\Edd}{\dot{M}^\RLOF_{\acc,i}}\right)\ ,
\label{eq:beta_rem}
\end{equation}
where the Eddington rate is given by $\dot{M}_\Edd = 4 \pi c R/\kappa_e$
with $\kappa_e$ the electron scattering opacity.

Another situation, referred to as an intermediate model by \cite{Huang1956} is when matter can escape the gravitational attraction of the binary system but does not leave to infinity. In this case, the expelled material can redistribute in a disk or ring surrounding the binary system. This is likely to happen if mass is lost through the \Ltwo{} point during common envelope evolution \citep{Ropke2023} or wind accretion \citep[e.g.][]{Chen2017}. Note also that in eccentric orbit or if the stars are not rotating synchronously, the deformation of the Roche potential gives favorable conditions for matter with moderate angular momentum to orbit the binary system (Sect.~\ref{sect:roche}). The specific angular momentum of the escaping material is then given by 
\begin{equation}
    j_\disc = [G(M_\donor+M_\gainer)\,a_\disc]^{1/2} \ ,
\end{equation}
where $a_\disc$ is the distance of the disk from the center of mass. The systemic angular momentum loss rate then writes 
\begin{equation}
    \dot{J}_\mass = j_\disc \dot{M} < 0 \ ,
\end{equation}
where $\dot{M}$ is given by Eq.~\ref{eq:mdot_sys}.
When the escaping material interacts with the binary environment, as illustrated in Fig.~\ref{fig:sph}, hydrodynamical simulations are necessary to obtain reliable estimates of the rate of angular momentum loss. From their SPH simulations of AGB binaries, \cite{Saladino2019} extracted a fitting expression for the angular momentum loss rate. They formulated their prescription in terms of an efficiency parameter $\eta$ which represents the fraction of the specific orbital angular momentum that is taken away by the wind so that
\begin{equation}
    \dot{J}_\mass = \eta \frac{J_\orb}{\mu} \dot{M}\ ,
\end{equation}
where $\mu = M_\donor M_\gainer/(M_\gainer+M_\donor)$ is the reduced mass and
\begin{equation}
\eta\left(q,\frac{v_\wind}{v_\orb}\right) = \min\left(\frac{1}{c_1+(c_2v_\wind/v_\orb)^3}+\eta_\iso,0.6\right)
\label{eq:def_eta}
\end{equation}
with $q = M_\donor/M_\gainer$, $c_1 = \max(q,\,0.6\, q^{1.7})$ and $c_2 = 1.5+0.3 q$. In the limit $v_\wind \gg v_\orb$, $\eta$ approaches the isotropic Jeans mode
\begin{equation}
\eta_\iso = \frac{1}{(1+q)^2} \ ,
\end{equation}
and with decreasing velocity ratio, the efficiency increases. In this regime, the wind interacts more strongly with the companion, enabling for a more efficient transfer of angular momentum. Simulations further indicate that $\eta$ increases as the mass ratio decreases, but it does not exceed 0.6. At low wind velocity, the spin velocity of the AGB star begins to affect $\eta$, but this effect remains moderate and is not included in their prescription. These results indicate that a substantial amount of angular momentum can  be extracted from the system, potentially leading to orbital shrinkage. This process can favor the occurrence of common envelope but also help reproducing the period distribution of Ba stars (see Sect.~\ref{sect:CBD}).

\subsection{Gravitational wave emission}
\label{sect:gw}

In close binary systems, especially those containing compact objects such as white dwarfs, neutron stars, or black holes, angular momentum loss due to gravitational wave radiation plays a significant role in driving orbital evolution. As the binary components orbit each other, they emit gravitational waves that carry energy and angular momentum away. This loss causes the orbital separation to decrease over time, leading to a gradual inspiral of the components. The rate of angular momentum loss is strongly dependent on the masses of the stars and the orbital separation. Its expression is given by \citep[e.g.][]{Tauris_vdH2023}
\begin{equation}
    \dot{J}_\gw = -J_\orb \frac{32 G^3}{5c^5}\frac{M_1M_2(M_1+M_2)}{a^4} \ .
\end{equation}
For systems with short orbital periods (typically less than a few hours), gravitational wave radiation can dominate over other mechanisms such as magnetic braking. They are considered to be the main driver of angular momentum loss in cataclysmic variables within the period gap (see Sect.~\ref{sect:magnetic}).

\subsection{Tidal interactions}
\label{sect:tides}

In close binary systems, tidal interactions arise when each star's gravitational field induces distortions in its companion, resulting in the formation of tidal bulges. Due to the finite viscosity and inertial lag in the stellar plasma, these bulges are offset by an angle $\alpha$ from the line connecting the stellar centers, creating a gravitational torque that transfers angular momentum between the stellar spins and the orbit. This process drives the system toward synchronous rotation and orbital circularization, as rotational kinetic energy is progressively converted into thermal energy.

A simple explanation of these effects can be deduced from energy considerations. The total energy of a binary system is given by
\begin{equation}
\mathcal{E} = \frac{1}{2}M_\donor v_\donor^{2}+\frac{1}{2}M_\gainer v_\gainer^{2}-\frac{GM_\donor M_\gainer}{r} = -\frac{GM\mu}{2a} \ , \label{eq:def_E}
\end{equation}
where $r$ is the separation between the two stars, $M=M_\donor+M_\gainer$, $\mu = M_\donor M_\gainer/(M_\donor+M_\gainer)$ is the reduced mass and $v_\donor,\ v_\gainer$ the orbital velocity of each star. If the orbital angular momentum and stellar masses are conserved, neglecting the stellar spins, the energy can be recast into
\begin{equation}
\mathcal{E}={\displaystyle - \frac{(GM)^2\mu^3}{2 J_\orb}\left(1-e^{2}\right)} \propto e^2-1 \ ,
\label{eq:Ebinary}
\end{equation}
leading to
\begin{equation}
\left(\frac{\partial\mathcal{E}}{\partial e}\right)_{J}\propto2e\,\,\,\mathrm{and}\,\,\,\left(\frac{\partial^{2}\mathcal{E}}{\partial e^{2}}\right)_{J}>0 \ .
\end{equation}
This expression shows that \emph{energy is minimum when $e=0$}, i.e. when the orbit has circularized.\\
If we now consider a circular orbit and a rotating star, the angular momentum and energy of the binary write
\begin{eqnarray}
J_\Sigma & = & I\Omega_\spin+\frac{M_{\donor}M_{\gainer}}{M}a^{2}\oorb \ ,\\ 
\mathcal{E} & = & \frac{1}{2}I\Omega_\spin^{2}-\frac{GM_{\donor}M_{\gainer}}{2a}\ .
\end{eqnarray}
Angular momentum conservation implies
\begin{equation}
\dot{J}_\Sigma=0\,\,\Rightarrow\,\,I\dot{\Omega}_\spin=\frac{1}{3}\frac{M_{\donor}M_{\gainer}}{M}a^{2}\dot{\Omega}_\orb \ ,
\end{equation}
where we used the Kepler law to have $-3\dot{a}/a=2\dot{\Omega}_\orb/\oorb$ and assumed that the star remains in full equilibrium ($\dot{I}=0$). Energy is minimum when $\dot{\mathcal{E}}=0$, i.e when
\begin{eqnarray*}
\frac{d\mathcal{E}}{dt} & = & I\Omega_\spin\dot{\Omega}_\spin+\frac{GM_{\donor}M_{\gainer}}{2a^{2}}\dot{a} \\ 
& = & I\Omega_\spin\dot{\Omega}_\spin-\frac{M_{\donor}M_{\gainer}}{3M}a^{2}\oorb\dot{\Omega}_\orb \\
& = & I\Omega_\spin\dot{\Omega}_\spin-I\oorb\dot{\Omega}_\spin=(\Omega_\spin-\oorb)I\dot{\Omega}_\spin\ .
\end{eqnarray*}
So the minimum of energy is found when the stars are synchronized, i.e. when $\Omega_\spin=\oorb$.

In stars, the dissipation of tides arises from the combination of (non-wave like) equilibrium tides and (wave-like) dynamical tides \citep{Zahn1977,Ogilvie2014}. Equilibrium tides are associated with the formation of tidal bulges and large-scale flows induced by the companion to maintain the star in hydrostatic equilibrium \citep{Zahn1966a,Remus2012}. Any process that resists this flow will lead to dissipation provided the viscosity is high enough. For this reason, equilibrium tides work efficiently in stars with convective envelopes, where viscous friction dissipates the kinetic energy of the tidal flow as it interacts with turbulent motions \citep{Zahn1977,Zahn1989}. 
\begin{figure}[t]
    \centering
    \includegraphics[width=\columnwidth]{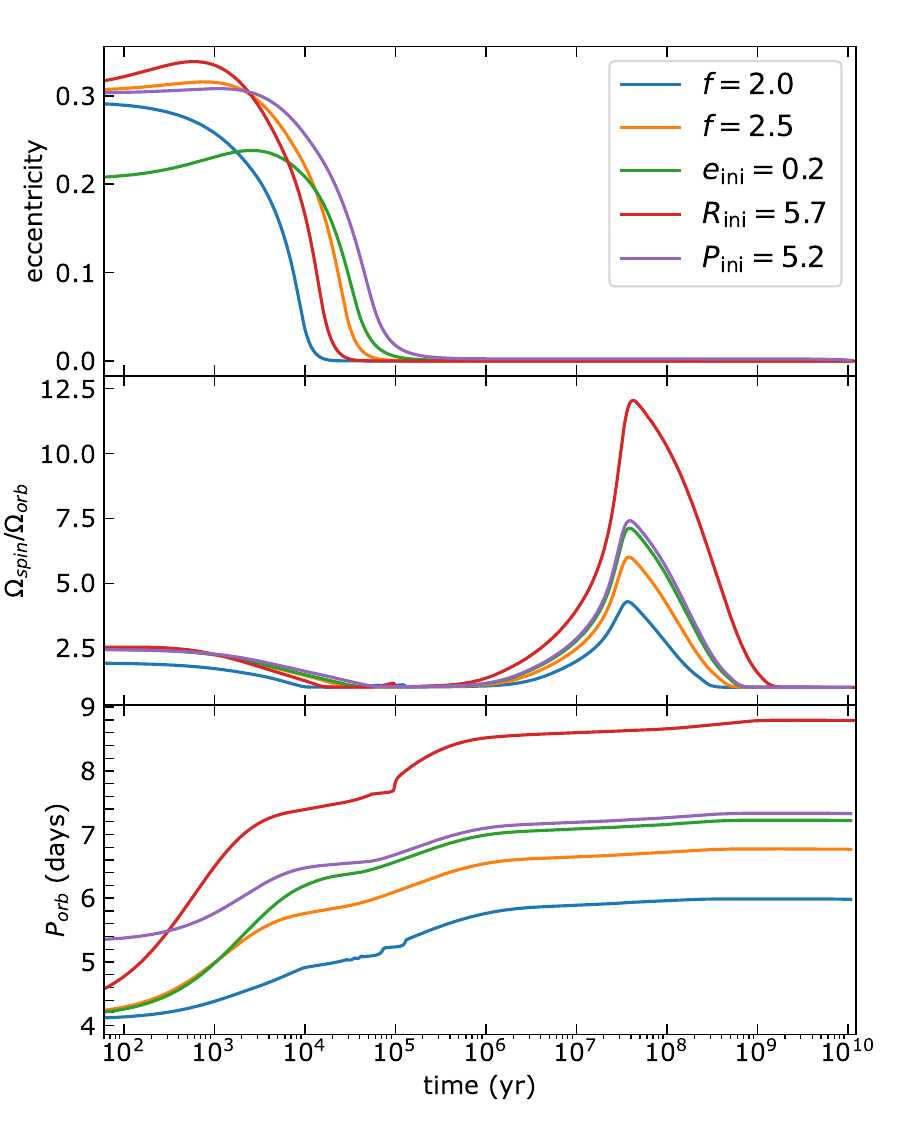}
    \caption{Evolution of the eccentricity (top), ratio of spin to orbital angular velocity (middle) and orbital period (bottom) during the pre-main sequence evolution of a 1.0+1.0\msun{} binary system. The default parameters are: initial period $P_\ini = 4$~d, eccentricity $e_\ini = 0.3$, radius $R_\ini = 4.7$\rsun, asynchronous factor $f=\Omega_\spin/\oorb = 2.5$. The reference track is shown in orange. }
    \label{fig:PMS_tides}
\end{figure}

Dynamical tides \citep{Zahn1977} arise from the excitation of internal waves near the convective core boundary that dissipate in the radiative envelope. This mechanism extracts energy from the oscillatory modes and contributes to the evolution of the orbital parameters and is more efficient than the energy dissipation by equilibrium tides in  stars with a convective core and a radiative envelope. Different types of waves can transport the energy and angular momentum in the radiative layers:
Acoustic waves  ($p$-modes)  are excited by the propagation of pressure perturbation but their frequency is generally too high to be excited directly by tidal forcing \citep{Ogilvie2014}. Inertial waves arise in rotating stars and have the Coriolis force as a restoring force. They have a relatively low frequency and require a high rotation rate to be effective \citep{Ogilvie2007}. In low- and intermediate-stars, dissipation of inertial waves is not expected to be a dominant process \citep{Esseldeurs2024}. Internal gravity waves ($g$-modes) have the buoyancy force as a restoring force. They respond to frequencies lower than the Brunt-Vaissala frequency and can be excited by the tidal potential. They emerge from the convective boundaries and dissipate their energy in the radiative layers near the surface where they can be partially reflected \citep{Zahn1975,Goldreich1989}.
An alternative approach to tidal dissipation was proposed by \cite{Tassoul1992a,Tassoul1992b} who developed a hydrodynamical model in which dissipation is very effective and occurs in an Ekman boundary layer near the stellar surface. This theory was contested by \cite{Rieutord1997} but a rebuttal was given by \cite{Tassoul1997}.

The efficiency of tidal dissipation processes depends strongly on the stellar structure (e.g. presence of convection zones), rotation, and fundamental parameters such as mass, metallicity or magnetic fields. Several expressions for the secular evolution of the orbital elements are available in the literature, but they depend on the dissipation mechanism considered.

For equilibrium tides, appropriate for stars with convective envelopes, the widely used equations of \cite{Hut1981} provide a general framework valid for any eccentricity $e$, unlike the earlier formulation by \cite{Zahn1977,Zahn1978} which was limited to small eccentricities. A refinement was later introduced by \cite{Zahn1989}, who modified his formalism to account for reduced dissipation when the tidal forcing frequency approaches or exceeds the convective turnover timescale (the so-called fast-tide regime).

For dynamical tides, the reference remains \cite{Zahn1977}, where the author derived expressions for the circularization ($\tau_\circ = |e/\dot{e}|$) and synchronization ($\tau_\sync=|\Omega_\orb-\Omega_\spin|/\dot{\Omega}_\spin$) timescales. Explicit expressions for the tidal torque ($\dot{J}^\tide_i = I_i\dot{\Omega}_\spin$) and circularization term ($\dot{e}$) are given in \cite{Hurley2002,Siess2013,Vos2015}. It should be noted however that using \cite{Hut1981} equations to treat dynamical tides can lead to inconsistencies and errors in the calculated stellar torque \citep{Sciarini2024}.

For equilibrium tides in convective stars, theory predicts that the timescales for orbital circularization and stellar synchronization scale as $\tau_\circ \propto (a/R)^8$ and $\tau_\sync \propto (a/R)^6$, respectively. Consequently, the ratio $\tau_\circ / \tau_\sync \propto (a/R)^2$ is greater than unity, indicating that corotation is usually achieved before the orbit becomes circular. This can be understood by noting that synchronization primarily affects the star, whose angular momentum is $J_\spin \approx M R^2 \Omega_\spin$, whereas circularization acts on the orbit, characterized by $J_\orb \propto \mu a^2 \oorb$, giving $\tau_\circ / \tau_\sync \propto J_\orb / J_\spin$.

Recent studies \citep[e.g.][]{Savonije1983,Savonije1984,Witte1999} suggest that tidal interaction can be significantly enhanced, at least for some time, if the star's natural oscillation frequencies (gravity modes) enter in resonance with the periodic tidal frequency in close binaries. In the regime of resonance locking, the evolution of both spin and orbit can be accelerated. This mechanism has been shown to operate in various astrophysical contexts, including white dwarf binaries \citep{Burkart2014}, heartbeat stars \citep[e.g.][]{Fuller2017}, and has even been proposed as a contributing factor in the orbital migration of exoplanets \citep[e.g.][]{Ma2021}.

Critical tests of tidal theories come from the determination of the circularization period, $P_\circ$, defined as the orbital period below which the observed binary systems (assumed to have roughly similar stellar masses and ages) are  circularized. \cite{ZahnBouchet1989} showed that circularization is very efficient during the pre-main sequence evolution of low-mass stars up to $P_\circ \sim 8$ days because young stars have extended convective envelope \citep[see also][]{Khaliullin2011}. Figure \ref{fig:PMS_tides} illustrates the rapid evolution of the orbital parameters for binary systems composed of two 1.0\msun{} stars under different initial conditions. An increase in eccentricity occurs in some models when the spin-to-orbital frequency ratio, $f = \Omega_\spin/\Omega_\orb$, exceeds $18/11$ \citep{Darwin1879}. Despite this, the systems circularize in less than a million years. The zero-age main sequence (ZAMS) is reached at approximately $10^8$~yr and coincides with the maximum value of $f$.  The final orbital period and the circularization timescale are determined by the initial conditions. Systems with higher angular momentum, resulting from faster spin rates, lower eccentricities, or longer initial periods, tend to circularize more slowly. In contrast, starting the evolution with a larger initial stellar radius enhances tidal interactions, leading to faster circularization.

Observationally, \cite{Meibom2005} analyzed samples of solar-type spectroscopic binaries of different ages.  In young clusters, they found that binaries are circularized below orbital periods of approximately 8 days (see their Figure 9), consistent with the predictions of \cite{ZahnBouchet1989}. Moreover, they observed an increase in the circularization period with age, reaching 15 days in the oldest clusters considered. The increase cannot be explained by equilibrium tides, whose efficiency on main sequence stars is too weak so another mechanism must operate to circularize the orbit, potentially associated with resonance locking. More recently, \cite{Bashi2023} analyzed a large sample of Gaia DR3 spectroscopic binaries including A-, F- and G-type main sequence primaries. They found that the circularization period depends weakly on stellar age but decreases linearly with increasing effective temperature, showing that hotter stars are circularized at shorter periods. This dependence on $T_\eff$ is not fully understood, but it likely traces the mass dependence of $P_\crit$. They conclude that pre-main sequence circularization is necessary to explain $P_\crit$, in line with the results of \cite{ZahnBouchet1989}. 
\cite{Verbunt1995} also find that the circularization period of binaries hosting a red giant star is in good agreement with \cite{Zahn1989} theory. This result was later confirmed by \cite{Price-Whelan2018} who used a larger sample of red giant binaries from the APOGEE survey. These observations also show an increase of $P_\crit$ with decreasing surface gravity, which is a proxy for age when the star evolves along the giant branch. As pointed out by \cite{Verbunt1995}, using giant stars removes some of the uncertainty of equilibrium tides because in such binaries, the orbital period is longer than the dissipation timescale associated with turbulent convection so the regime of fast tides is avoided. For more massive stars with a radiative envelope where the dynamical tides are most efficient, observations of eclipsing binaries in the small and large Magellanic clouds \citep{North2003} also tend to validate the theory devised by \cite{Zahn1975}. However, a large fraction of systems are also circularized at periods larger than the predicted transition period. This has been confirmed in recent TESS survey \citep{Justesen2021} suggesting that additional mechanisms are operating in stars with radiative envelope. Potential candidates involve resonance locking \citep{Witte1999} or inertial waves, but currently there is no consensus on a model that could explain the circularization of long-period systems \cite[see reviews by][]{Mazeh2008,Zahn2008}.

Tides also act to synchronize the stellar spins with the orbital period so in principle a method similar to that used for the circularization can be used. This exercise is however more challenging because it can be difficult to determine the stellar rotation period if the inclination is badly constrained and also because the synchronization of the stellar surface may hide the fact that its interior is in differential rotation \citep[for a review, see][]{Mazeh2008}.

\subsection{Eccentric orbits}
\label{sect:eccentric}

In eccentric orbits, variations in stellar separation and orbital angular velocity throughout the orbital cycle affect the rates at which mass and angular momentum are transferred or lost. These changes can, in turn, modify the system's eccentricity. The theoretical framework describing these effects is presented below.

\subsubsection{Eccentricity generation terms due to mass exchange}
\label{sect:edots}

To estimate the variation in eccentricity resulting from the exchange of mass between binary components, we adopt the elegant approach presented by \cite{Eggleton_book}, which is based on the Laplace-Runge-Lenz (LRL) vector, whose expression is given by
\begin{equation}
GM\mathbf{e}=\mathbf{d}\times\mathbf{\dot{h}}-\frac{GM}{r}\mathbf{r}\ ,
\end{equation}
where $M=M_{\donor}+M_{\gainer}$, $\mathbf{r}=\mathbf{r}_{\donor}-\mathbf{r}_{\gainer}$ is the (instantaneous) distance between the 2 stars and $\mathbf{\dot{h}}=\mathbf{r} \times \mathbf{\dot{r}}$ the specific orbital angular momentum. The vector $\mathbf{e}$ has very interesting properties : it points toward the periastron and has a modulus equal to the eccentricity.
The rate of change of the stellar masses can, in a general form, be expressed as
\begin{eqnarray*}
\dot{M}_{\donor} & = &  \dot{M}_\donor^\loss-|\Mtrans| \ ,\\
\dot{M}_{\gainer} & = & \dot{M}_2^\loss+|\Mtrans| \ ,
\end{eqnarray*}
where $\dot{M}_i^\loss<0$ is the net rate of mass loss to infinity from star $i$ and $\Mtrans$ represents the mass transfer rate between the two stars either by RLOF or wind accretion. In this framework, the mass lost by the system is given by $\dot{M}=\dot{M}_\donor^\loss+\dot{M}_\gainer^\loss$. With this parameterization and  {\em assuming mass is lost isotropically} so it does not affect the motion of the stars, one finds 
\begin{equation}
\mathbf{\dot{e}}=\left(2\Mtrans\left[\frac{1}{M_{\donor}}-\frac{1}{M_{\gainer}}\right]-\frac{\dot{M}}{M}\right)\left[\begin{array}{c}
\cos\nu+e\\
\sin\nu\\
0
\end{array}\right]\label{eq:edot_general}
\end{equation}
where $\nu$ is the true anomaly. 
In Eq.~\ref{eq:edot_general}, the change in eccentricity is associated with the radial component of the LRL vector, the tangential component being related to the apsidal motion (precession of the orbit). In cases where mass transfer results from both wind accretion ($\dot{M}^\wind_{\acc,3-i} = -\beta_i \dot{M}^\wind_i$) and non-conservative Roche lobe overflow from star 1 to star 2 ($\dot{M}^\RLOF_{\acc,2}=-\beta_\RLOF \dot{M}^\RLOF$), the expressions for $\dot{M}_i^\loss$ and $\Mtrans$ can be written as follows
\begin{eqnarray}
 \dot{M}_1^\loss & = & (1-\beta_\donor)\dot{M}_\donor^{\wind}+(1-\beta_\RLOF)\dot{M}^{\RLOF} \ ,\\
    \dot{M}_2^\loss & = & (1-\beta_\gainer)\dot{M}_\gainer^{\wind} \ , \\
    \Mtrans & = & -\beta_\RLOF \dot{M}^{\RLOF} -  \beta_1 \dot{M}_\donor^{\wind} +\beta_2\dot{M}^{\wind}_2 \ ,
\end{eqnarray}
from which we obtain
\begin{eqnarray}
\dot{e} = &  {\displaystyle \frac{-(1-\beta_\donor)\dot{M}_\donor^{\wind}-(1-\beta_\gainer)\dot{M}_\gainer^{\wind}-\left(1-\beta_\RLOF\right)\dot{M}^{\RLOF}}{M}} \nonumber \\
& \times \left(e+\cos\nu\right) \nonumber\\
 & {\displaystyle -2\left[\beta_\RLOF\dot{M}^{\RLOF}+\beta_\donor\dot{M}_\donor^{\wind}-\beta_\gainer\dot{M}_\gainer^{\wind}\right]\left(\frac{1}{M_\donor}-\frac{1}{M_\gainer}\right)} \nonumber \\ 
& \times \left(e+\cos\nu\right) \ . 
\end{eqnarray}
From this expression, various contributions can be extracted depending on whether mass is lost, accreted or transferred by wind or RLOF :
\begin{eqnarray}
\dot{e}_{\loss,i}^{\wind} & = & {\displaystyle -\frac{\dot{M}_{i}^{\wind}}{M}\left(e+\cos\nu\right)}\label{eq:edot_wind_loss}\ ,\\
\dot{e}_{\acc,i}^{\wind} & = & {\displaystyle -\frac{\dot{M}_{\acc,i}^\wind}{M}\left(e+\cos\nu\right)}\label{eq:edot_wind_acc}\ ,\\
\dot{e}_{i}^{\wind,\mathrm{ex}} & = & {\displaystyle -2\,\dot{M}^\wind_{\acc,i}\left(\frac{1}{M_{i}}-\frac{1}{M_{3-i}}\right)\left(e+\cos\nu\right)} \label{eq:edot_wind_ex} \ ,\\
\dot{e}_\donor^{\RLOF} & = & {\displaystyle -\left(1-\beta\right)\frac{\dot{M}^{\RLOF}}{M}\left(e+\cos\nu\right)}\label{eq:edot_RLOF_loss}\ ,\\
\dot{e}_\gainer^{\RLOF} & = & {\displaystyle -2\beta\left(\frac{1}{M_\donor}-\frac{1}{M_\gainer}\right)\dot{M}^{\RLOF}\left(e+\cos\nu\right)} \label{eq:edot_RLOF_acc}\ , \\
\dot{e}_i & = & \left(\dot{e}_{\loss,i}^{\wind}+\dot{e}_{\acc,i}^{\wind}+\dot{e}_{i}^{\RLOF}+\dot{e}_{i}^{\wind,\mathrm{ex}}\right)\ .
\label{eq:edot_i}
\end{eqnarray}\\
The term $\dot{e}_{i}^{\wind,\mathrm{ex}}$ corresponds to the change in eccentricity as a result of wind mass transfer.  Note that if both $\Mtrans$ and $\dot{M}$ are phase independent, then
\[
\langle\dot{e_i}\rangle=\frac{1}{P}\int_{0}^{P}\dot{e}_i\,dt=\frac{\left(1-e^{2}\right)^{3/2}}{2\pi}\int_{0}^{2\pi}\frac{\dot{e_i}\,d\nu}{\left(1+e\cos\nu\right)^{2}}=0\ ,
\]
implying that mass exchange between the stellar components or mass loss through isotropic ejection (Jeans mode) does not affect the eccentricity.

A common simplification often adopted in analytical treatments is to assume that both mass-loss and mass-accretion occur isotropically, and that mass transfer does not involve any exchange of linear momentum. These assumptions are, however, inadequate for modeling situations like Roche‑lobe overflow, where the flow geometry is inherently anisotropic. More refined models have incorporated the momentum exchange associated with directional mass transfer.
Several approaches have been developed to address this problem, treating the effect of mass transfer as a perturbation of the instantaneous (osculating) orbital elements \citep[e.g.][]{Sepinsky2009,Dosopoulou2016,Rocha2025,Parkosidis2026a} or by considering changes in the total binary orbital energy and angular momentum \citep[e.g.][]{Huang1956,Bonavic2008}. 
These improved formalisms have yielded secular prescriptions for the evolution of orbital parameters, applicable to both circular and eccentric binaries, conservative and non‑conservative mass‑transfer regimes.

From the above expressions (Eq.~\ref{eq:edot_wind_loss}-\ref{eq:edot_RLOF_acc}), it becomes evident that phase-dependent mass transfer rates can modify the eccentricity. Such variations may be induced by Roche lobe overflow occurring preferentially at periastron, from tidal interactions (e.g., Eq.~\ref{eq:CRAP}), from changes in wind-focusing efficiency (Eq.~\ref{eq:beta_BH}), or from intrinsic fluctuations in the stellar wind mass-loss rate driven by convection, pulsations, or rapid structural evolution.

Several studies have investigated the role of phase-dependent wind mass loss in shaping the orbital evolution of systems containing a RGB or AGB star \citep[e.g.][]{Soker2000, Bonavic2008}. In particular, this mechanism has been invoked 
to explain the high eccentricities observed in companions to evolved giants and their descendants, including hot subdwarfs \citep{Vos2015}, giant barium stars \citep{Escorza2020,Krynski2025} or helium white dwarfs  \citep{Siess2014}. These works show that phase dependent wind mass loss can enhance orbital eccentricity, partially offsetting the circularizing effects of tides. Systems such as IP~Eri \citep{Siess2014} and Sirius \citep{Bonavic2008}, which exhibit eccentric orbits despite significant tidal interaction, can be naturally explained within this framework. 

\begin{figure}
    \centering
    \includegraphics[width=\columnwidth]{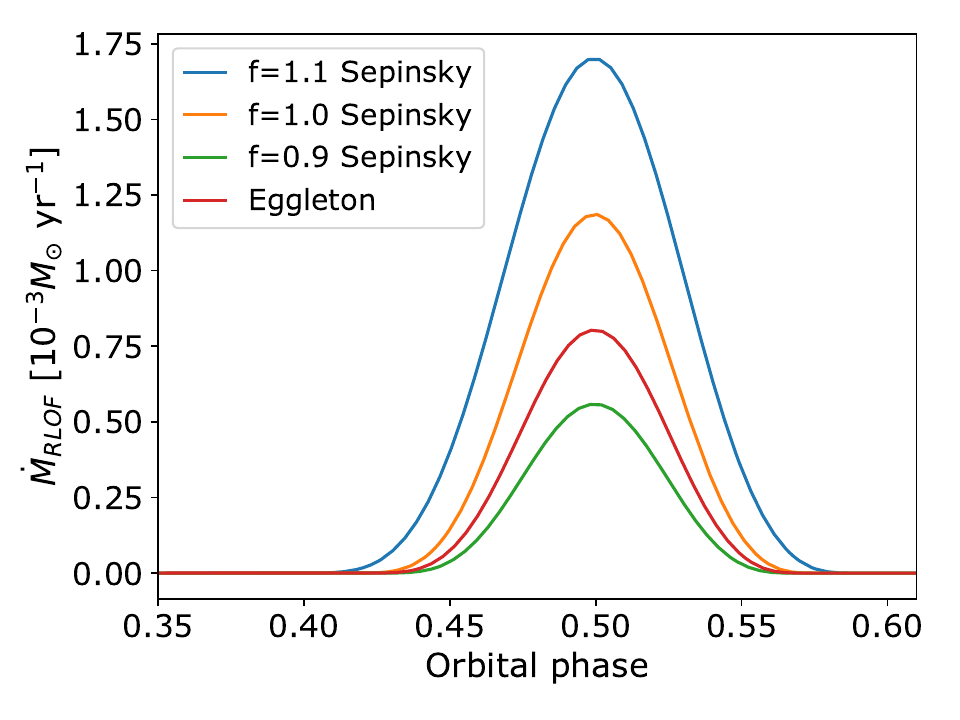}
    \caption{Roche lobe overflow mass transfer rate as a function of orbital phase for different prescription for the Roche radius and values of the asynchronicity factor $f$. The binary system consists of a 1.5\msun{} donor and a 1.4\msun{} accretor with an orbital period of 6~days and an eccentricity $e=0.25$. The mass transfer rate is computed using the \cite{Kolb1990} formalism, adopting either the Roche radius expression from \cite{Eggleton1983} which is independent of $f$ (blue line) or the formulation of \cite{Sepinsky2007b} derived from Eq.~\ref{eq:phi_sepinsky}, with $f = 0.9$, 1, and 1.1. Periastron occurs at orbital phase 0.5. }
    \label{fig:mdot_phase}
\end{figure}

Roche‑lobe overflow in eccentric binaries has been examined in detail by \citet{Davis2013}. Figure~\ref{fig:mdot_phase} illustrates the resulting phase‑dependent mass‑transfer rates for a system composed of a 1.5\msun{} donor and a 1.4\msun{} companion in a 6‑day orbit with eccentricity $e = 0.25$. As expected, mass transfer proceeds episodically near periastron, where the orbital separation reaches its minimum. The resulting mass‑loss episodes exhibits a Gaussian‑like profile, in agreement with the smoothed‑particle hydrodynamics simulations of \citet{Lajoie2011}.
The magnitude of the mass‑transfer rate is sensitive to the adopted prescription for the Roche‑lobe radius. In the formalism of \citet{Sepinsky2007b}, which explicitly incorporates the effects of orbital eccentricity and spin asynchronicity, increasing the rotation factor $f$ (Eq.~\ref{eq:def_f}) reduces the Roche‑lobe radius and correspondingly enhances the mass‑transfer rate. In contrast, applying the standard \citet{Eggleton1983} approximation (Eq.~\ref{eq:Roche_radius}) yields mass‑transfer rates that are lower relative to the synchronous case ($f = 1$) and leads to a shorter semi‑detached phase.

\subsubsection{Secular evolution}
Evolutionary timescales can be orders of magnitude longer than the orbital period. In eccentric orbits, the calculation of phase-dependent quantities, such as the stellar torques and mass transfer rates, requires a specific treatment. One possible approach, albeit time-consuming, is to resolve the orbit using small timesteps ($\Delta t/P_\orb \ll 1$) to follow the evolution of the orbital parameters. An alternative is to use analytical expressions for the averaged quantities. Such prescriptions have been provided by \cite{Sepinsky2007b,Sepinsky2009} and \cite{Dosopoulou2016} but in these derivations, the mass loss rate is modeled by a delta function centered at periastron. This choice is valid for high eccentricity or almost circular orbits but provides a crude estimate outside these two specific cases. Improved formulations that relax the assumption of instantaneous mass transfer at periastron have recently been proposed by \cite{Hamers2019,Parkosidis2026a,Parkosidis2026b} but these prescriptions have not yet been incorporated in binary star evolution calculations. A general numerical solution, independent of the expression for the mass transfer rate, involves the use of Gaussian quadrature\footnote{\url{http://en.wikipedia.org/wiki/Gaussian\_quadrature}} \citep{Siess2014}. The technique approximates an integral by a weighted sum of the function under consideration ($f$), evaluated at specific points in the integration domain, i.e.
\begin{equation}
  \int_{a}^{b}f(t)\,dt\approx\frac{b-a}{2}\sum_{i=1}^{n}w_{i}f\left(\frac{b-a}{2}x_{i}+\frac{a+b}{2}\right)\ ,
  \label{eq:gauss}
\end{equation}
where $w_i$ are weight factors and $x_i$ specific positions. The parameters $w_i$ and $x_i$ are tabulated quantities\footnote{\url{https://pomax.github.io/bezierinfo/legendre-gauss.html}} that depend on the order $n$ of the sum. In an eccentric orbit, the average of a quantity $f$ over the time interval $[\tau_{0},\tau_{1}]$ is equivalent to averaging over the corresponding true anomaly interval $[\nu_{0},\nu_{1}]$, and is given by
\begin{eqnarray*}
\langle f \rangle & = & \frac{1}{\tau_1-\tau_0}\int_{\tau_{0}}^{\tau_{1}}f(t)dt\ =\ \frac{1}{\tau_1-\tau_0}\int_{\nu_{0}}^{\nu_{1}}f(\nu)\frac{d\nu}{\omega(\nu)}\\
 & = & \frac{\left(1-e^{2}\right)^{3/2}}{2\pi}\frac{P_\orb}{\tau_1-\tau_0} \int_{\nu_{0}}^{\nu_{1}}\frac{f(\nu)}{\left(1+e\cos\nu\right)^{2}}d\nu\ .
\end{eqnarray*}
Defining the function $g(\nu) = f(\nu)/(1+e\cos\nu)^{2}$, the orbital average reads
\begin{equation}
\int_{\nu_{0}}^{\nu_{1}} f(\nu)\,d\nu\ \approx\ \frac{\nu_{1}-\nu_{0}}{2}\sum_{i=1}^{n}w_{i}\,g\left(\frac{\nu_{1}-\nu_{0}}{2}x_{i}+\frac{\nu_{1}+\nu_{0}}{2}\right)\ .
\end{equation}
In practice, all phase-dependent quantities should be averaged. This includes the mass transfer rates $\dot{M}_i$, possibly the wind mass loss rate in case it is tidally enhanced (Eq.~\ref{eq:CRAP}), the rates of change of the eccentricity (Eq.~\ref{eq:edot_i}) and the stellar torques.

\subsection{Circumbinary discs}
\label{sect:CBD}

Circumbinary disks (CBDs) have been detected in a wide range of systems, including young stars \cite[e.g.,][]{Dutrey2016,Cuello2025}, evolved AGB binaries \citep{Kervella2016}, and all known post-AGB binaries \citep{Winckel2018}. Their presence is typically inferred from near-infrared excess emission in the spectral energy distribution, which signals the existence of warm dust. In post-AGB binaries, CBD masses are estimated to range between $10^{-3}$ and $3\times10^{-2}\Msun$ \citep{Bujarrabal2013}. Radiation hydrodynamical simulations of young stellar clusters yield comparable CBD masses for protostellar systems, spanning $10^{-4}$ to $10^{-1}\Msun$ \citep{Elsender2023}.

In evolved stars, circumbinary disk formation is attributed to mass loss through the outer Lagrangian points (\Ltwo{} or \Lthree), either as a consequence of Roche-lobe overflow \citep{Pejcha2016}, wind-driven mass transfer \citep{Chen2017}, or common envelope evolution \citep{Ropke2023}. Material escaping through the outer Lagrangian point is shaped by orbital motion into a spiral structure that wraps around the system. If the binary is sufficiently massive, this material remains gravitationally bound and eventually settles into a CBD \citep{Frankowski2007,MacLeod2018}.

The stars' orbital motion induces a varying gravitational potential, exciting spiral density waves in the circumbinary disc. These waves transport energy and angular momentum from the binary to the disk \citep{Goldreich1978, Goldreich1979}, thereby influencing the orbital evolution of the system. The interaction is strongest at the Lindblad resonances, where the natural frequencies of disk particles resonate with the binary's orbital motion \citep{Artymowicz1991}. These resonances inhibit accretion onto the binary, creating a low density central cavity \citep{Pringle1991}. The size of the cavity  depends on the binary parameters (separation, mass ratio, eccentricity) and disk properties (viscosity, scale height). In circular orbits, the radius of the cavity is typically $\sim 2-3$ times the orbital separation, while eccentric binaries generate larger cavities, scaling approximately as $(1+e)$ \citep{Artymowicz1994}.

Despite the presence of this tidal barrier, material can still cross the cavity, as seen in the T Tauri binary system GG Tau \citep{Dutrey2016}. Indirect evidence for mass exchange between the CBD and the central stars comes from the depletion of refractory elements in post-AGB binary companions \citep{Winckel1995}. This chemical anomaly is interpreted as the reaccretion of metal-poor gas from the CBD, where refractory elements have condensed into dust and been removed from the gas phase. Hydrodynamical simulations further confirm that gas can penetrate the cavity \citep[e.g.][]{Artymowicz1996}, often in the form of intermittent burst-like episodes \citep{Munoz2016}.

The CBD can significantly affect the binary's orbital evolution. In the regime of low eccentricities ($e < 0.2$), \cite{Lubow1996} derived the following expressions for the changes in separation and eccentricity by the resonant interactions 
\begin{eqnarray}
  \left(\frac{\dot{a}}{a}\right)_\res & = & \displaystyle - \frac{2l}{m} \frac{\dot{J}_\res}{J_\orb} \ , \label{eq:adot_res} \\
  \dot{e}_\res  & = &  \displaystyle - \left(\frac{1-e^2}{e}\right) \left(\frac{m}{2l}\right)\left[\frac{l}{m}-\frac{1}{(1-e^2)^{1/2}}\right]\left(\frac{\dot{a}}{a}\right)_\res\ , 
  \label{eq:edot_res}
\end{eqnarray}
where $\dot{J}_\res$ is the resonant torque exerted on the orbit by the disk and $(m,l)$  the excited potential harmonics. At low eccentricity, \cite{Artymowicz1994} showed that the dominant interaction occurs at the outer Lindblad resonance (LR), corresponding to the $(m,l)=(2,1)$. For higher eccentricities, additional resonances are excited and the eccentricity pumping is significantly damped.
The torque exerted on the disk at the LRs ($\dot{J}_\disc$) was determined by \cite{Goldreich1978,Goldreich1979} using linear perturbation theory and can be written as
\begin{equation}
  \dot{J}_\disc= -\dot{J}_\res = -m\pi^{2}\Sigma(r_\res)|\Psi|^{2}\left(r\frac{\mathrm{d}\mathcal{D}}{\mathrm{d}r}\right)^{-1},
\label{eq:Jdot_disc}
\end{equation}
where $\Sigma(r_\res)$ is the disc's surface density at the LR, $\mathcal{D}$ a measure of the distance from the LR, and $\Psi$ is the forcing function (the details of these expressions are not needed here but can be found in the previous references). At the outer LR, which is most relevant for a CBD, $\dot{J}_\disc >0$ indicating that the disk extracts angular momentum from the orbit. For nearly Keplerian CBD, \cite{Krynski2025} showed that in the limit of small $e$ when the $(m,l)=(2,1)$ resonance dominates, the resonant torque takes the form
\begin{equation}
  \dot{J}_\res= - \frac{49 \pi^2}{8} G\, \frac{\mu^2}{M} a\,  \Sigma(r_\res)\, e^2,
\label{eq:Jdot_ml}
\end{equation}
where $\mu=M_{\donor}M_{\gainer}/(M_{\donor}+M_{\gainer})$ is the reduced mass and $M=M_\donor+M_\gainer$. 

An alternative expression for $\dot{J}_\res$ was proposed by \cite{Lubow1996} based on 2D SPH simulations. These authors found that the torque on the disk is relatively independent of the strength and radial extent of the resonance area and can be approximated as
\begin{equation}
    \dot{J}_\disc \approx \frac{J_\disc}{\tau_\nu} = J_\disc \,\alpha \left(\frac{H}{R}\right)^2 \Omega_\disc= M_\disc \Omega_\disc^2 R^2 \alpha \left(\frac{H}{R}\right)^2\ ,
    \label{eq:Jdot_disc_approx}
\end{equation}
where $J_\disc$ is the angular momentum of the disc, $\tau_\nu= R^2/\nu$ the disk viscous timescale, $\nu$ the viscosity, $\alpha$ the viscosity parameter \citep{Shakura1973} and $M_\disc$ the mass of the disc. The (Keplerian) disk angular velocity $\Omega_\disc$ is evaluated at a characteristic disk radius, typically associated with the half-angular momentum radius. The two approaches for modeling the resonant torque in CBDs depend on distinct parameters: the surface density at the Lindblad resonance (LR) in one case, and global disk properties (such as mass and radial extent) in the other. At very low eccentricities, substituting Eq.~\ref{eq:Jdot_ml} into Eq.~\ref{eq:edot_res} with the use of Eq.~\ref{eq:edot_res} yields $\dot{e}_\res \propto e$, consistent with results from smoothed particle hydrodynamics (SPH) simulations \citep{Lubow1996}. However, this linear proportionality does not hold when using the formulation given by Eq.~\ref{eq:Jdot_disc_approx}.

The expression for $\dot{e}_\res$ (Eq.~\ref{eq:edot_res}) must also be adjusted to account for the reduction of the tidal torque at high eccentricities when more resonances are excited. \citet{Lubow1996} showed that the dependence of $\dot{e}_\res$ on $e$ follows three regimes: a linear rise up to $e \simeq 0.1\alpha^{1/2}$, a decline approximately proportional to $1/e$ in the intermediate range ($0.1\alpha^{1/2} \lesssim e \lesssim 0.2$), and finally saturation to values close to zero at high eccentricity. Simulations of black hole binaries with $e \approx 0.6-0.8$ \citep{Roedig2011} confirm the eccentricity saturation in this regime, with $\dot{e}_\res \approx 0$.  Similar behavior is found in 2D SPH simulations of equal-mass binaries by \cite{Dorazio2021}. To model this transition, \cite{Dermine2013} and \cite{Krynski2025} proposed a simple solution, consisting of multiplying Eq.~\ref{eq:edot_res} by some ad-hoc function.  

Numerical simulations by \cite{Munoz2019}, focusing on equal-mass binaries, reveal that mass accretion from the CBD leads to a net gain of angular momentum by the binary at an average rate of
\begin{equation}
    \dot{J}_\acc \approx 0.7 \dot{M}_\bin\, a^2 \oorb \ ,
    \label{eq:Jacc_disc}
\end{equation}
where $\dot{M}_\bin$ is the mass accretion rate on the binary. This rate appears to be relatively independent of the binary mass ratio \citep{Lai2023}. 
The influence of this accretion stream on the eccentricity evolution is less straightforward. Analytical models by \citet{Rafikov2016} suggest that sustained eccentricity growth in post-main-sequence binaries requires a relatively massive ($M_\disc \gtrsim 10^{-2}\Msun$), long-lived ($\gtrsim 10^5$ yr) CBD that does not undergo significant re-accretion. By contrast, the simulations of \citet{Munoz2019} display a non-monotonic behavior, with $\dot{e} \approx {0, 2.42,-0.2,-2.34}$ for initial eccentricities $e={0,0.1,0.5,0.6}$, respectively. This puzzling behavior may stem from limitations in the resolution of the simulations, thermal effects which have been shown to influence orbital dynamics \citep{Wang2023} or related to bifurcation in CBD precession rates.  The role of reaccretion in modulating or suppressing eccentricity pumping therefore remains uncertain and merits further investigation.

Binary evolution models have begun to incorporate the physics of CBD into simulations of interacting binaries. Using a population synthesis code, \cite{Dermine2013} investigated the effect of the CBD on the orbital evolution of post-AGB stars. Their simulations, based on the formalism outlined in Eqs.~\ref{eq:edot_res} and \ref{eq:Jdot_disc_approx},  showed that eccentricity can be efficiently excited, supporting the idea that CBDs provide a viable mechanism for increasing orbital eccentricity. Building on this framework, \cite{Vos2015} implemented this formalism in the MESA binary code to study the formation of long-period hot subdwarf-B (SdB) binaries. These compact core-He burning stars have lost their H-rich envelope and display increasing eccentricities between 0 and 0.25 in the period range $750-1350$~days at odds with tidal models that predict circularized systems. To explain this discrepancy, the authors explored several mechanisms capable of driving orbital eccentricity, including those related to mass transfer (see Sect.~\ref{sect:eccentric}) and CBD physics. In the later approach, they adopted a parametric surface density profile $\Sigma(r)$ to estimate the disk angular momentum in Eq.~\ref{eq:Jdot_disc_approx}. However, their models yield a trend opposite to that observed, predicting higher eccentricities at shorter orbital periods.

\begin{figure}
    \centering
    \includegraphics[width=\columnwidth]{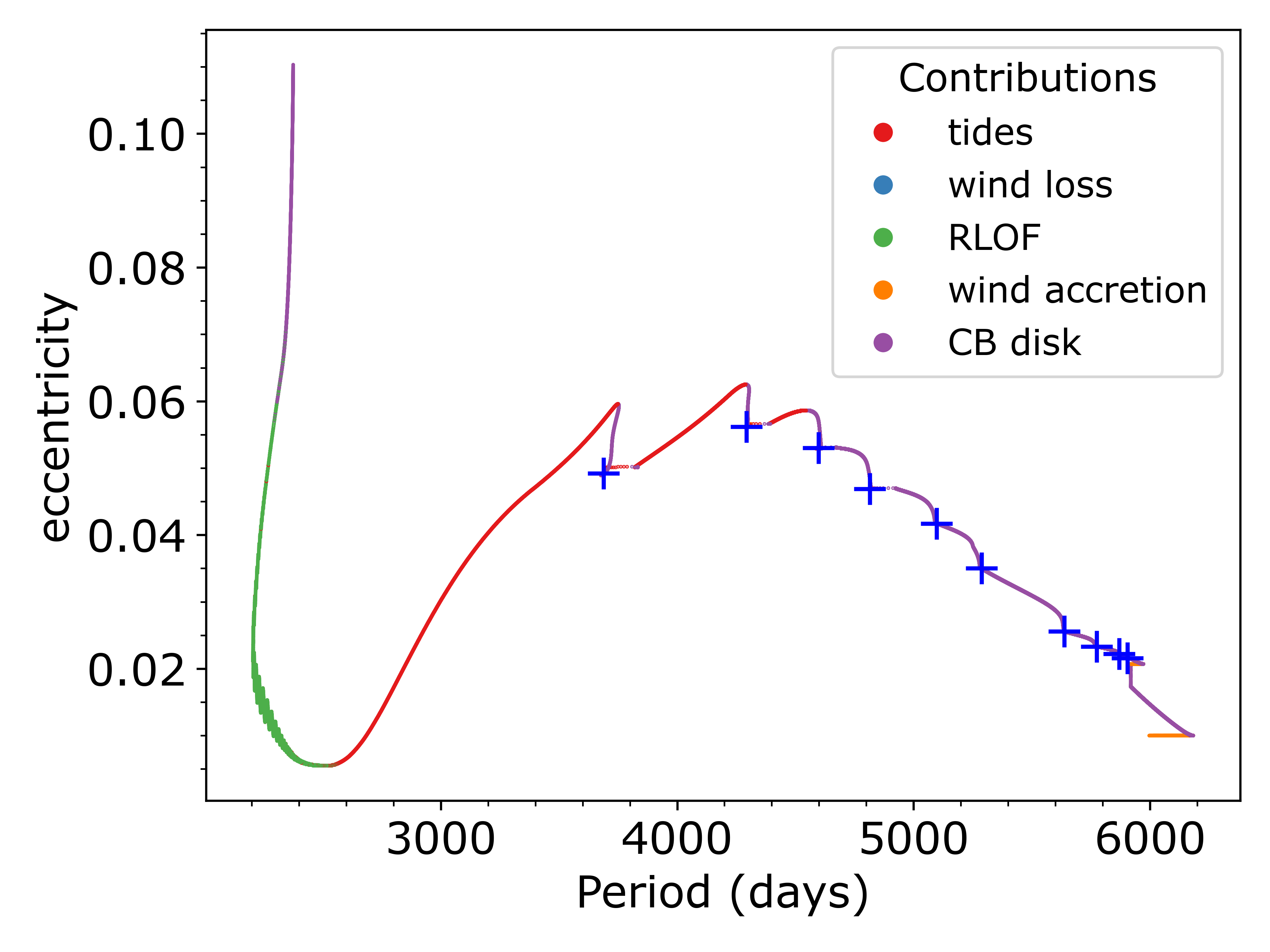}
    \caption{Evolution in the eccentricity-period diagram of a 2+1\msun{} binary system with initial period $P_\orb=6000$ days and eccentricity $e = 0.01$. The evolution is non conservative with $\beta_\RLOF = 0.5$. The colors on the curve display the different processes that affect the eccentricity. The blue crosses mark the occurrence of thermal pulses. }
    \label{fig:elogP_CBD}
\end{figure}

More recently, \cite{Krynski2025} analyzed the effect of a CBD on the formation of Ba stars, which, similarly to SdB binaries, exhibit significant eccentricities in a period range where tidal forces are expected to efficiently circularize the orbit. One of their simulations, shown in Fig.~\ref{fig:elogP_CBD}, illustrates the effects of various processes in shaping the orbital evolution. The system begins as a 2+1\msun{} binary with a period of 6000 days and an eccentricity $e=0.1$. Up to the beginning of the AGB phase, the separation increases due to Jeans-mode wind mass loss, while the eccentricity remains nearly constant because tides are inefficient at such large separations. Once the primary reaches the AGB, the conditions for WRLOF become favorable and the associated angular momentum loss prescription (Eq.~\ref{eq:beta_BH}) leads to orbital contraction. The feeding of the CBD during the phase of WRLOF activates the eccentricity pumping. Below $P_\orb \lesssim 4500$~d, tides begin to circularize the orbit but during the thermal pulses (blue crosses), the CB pumping (Eq.~\ref{eq:edot_res}) momentarily dominates, producing the saw tooth pattern in the $e-\log P$ diagram. Near the AGB tip, because of the large stellar radius and considerable reduction of the orbital separation, the system circularizes. At $P_\orb \approx 2700$~days, the AGB star fills its Roche lobe and mass transfer starts. At this stage the primary is less massive than the secondary. This stabilizes mass transfer (Eq.~\ref{eq:adot_conservative}) and generates some eccentricity (Eqs.~\ref{eq:edot_RLOF_loss} and \ref{eq:edot_RLOF_acc}). Eventually the system detaches and the CBD disk pumping takes over, until it dissipates. In the end the system reaches a period of $\sim 2660$~days and an eccentricity $e \approx 0.11$. This complex modeling helps in reproducing the observed orbital properties of Ba stars, however it still fails to account for the population of eccentric binaries with orbital periods shorter than $\sim 2000$~days.  An interesting result of their simulations is also the occurrence of Roche-lobe overflow from AGB stars in eccentric orbits.

\section{Discussion}
\label{sect:discussion}

The evolution of binary systems relies on a complex interplay between processes that control how mass and angular momentum are exchanged between the stellar components and with the outer binary environment. While progress has been made, substantial uncertainties persist in our 1D binary evolution models. To conclude this review, we briefly highlight some areas where better understanding is needed.

Our traditional reliance on Roche-lobe overflow models, rooted in 1D approximations, fails to capture the multidimensional complexity revealed by modern simulations \citep[e.g.][]{Scherbak2025}, particularly in eccentric, asynchronously rotating systems. Hydrodynamical simulations are therefore essential to refine the prescriptions implemented in 1D stellar evolution codes, but they must also account for additional complexities, such as mass transfer in eccentric binaries \citep{Lajoie2011,Staff2016} and quantify the spin-up of the accretor.

Concerning wind mass transfer, current prescriptions \citep{Abate2013,Saladino2019} are based on hydrodynamical models that probe a restricted region of the parameter space and adopt simplified treatments of the underlying physics. In particular, the lack of a self-consistent description of dust-driven winds, including chemistry, dust formation, shocks, cooling, and radiative transfer, limits the robustness of the derived prescriptions \citep[e.g.][]{Maes2022}.

As discussed in Sect.~\ref{sect:CBD}, a CBD can significantly affect orbital evolution through resonant interactions. Recent studies \citep{Dorazio2021,Zrake2021,Siwek2023,Valli2024,Murray2025} have shown that eccentricity in CBD–binary systems tends to reach an equilibrium value around $e \approx 0.4$–0.6, set by the balance between resonant pumping and damping from disk viscosity and re-accretion. These findings challenge earlier analytical models \citep[e.g.][]{Lubow1996} and underscore the need for advanced simulations to capture these effects. 
In a recent study, \cite{Huang2025} showed that in post-AGB binaries, interactions with misaligned CBDs can drive eccentricity pumping. In this configuration, the torque between the misaligned disk and the binary produces periodic variations in inclination and eccentricity, known as von Zeipel–Kozai–Lidov (ZKL) oscillations.
The efficiency of eccentricity growth depends strongly on the degree of misalignment, with larger tilt angles producing higher eccentricities. Once the disk dissipates, the ZKL oscillations cease, potentially leaving the binary in a long-lived high-eccentricity state, as observed in many post-AGB systems. This mechanism offers an attractive explanation for the long-standing $e$–$\log P$ problem in post-AGB binaries, and may also be relevant for understanding the puzzling orbital properties of Ba stars as well as CEMP, CH and symbiotic stars. Observationally, however, constraints on CBD properties such as their masses and orientations remain scarce, though new facilities like ALMA,  JWST and upcoming ELT will provide critical insights on these influential structures.

Tidal dissipation is another key process shaping the orbital and rotational evolution of close binary stars and planetary systems, yet its fundamental physics remains incompletely understood.Recent numerical studies, including 3D magnetohydrodynamical simulations, have begun to address some of these gaps by exploring how tides interact with magnetic fields \citep[e.g.,][]{Astoul2022,Astoul2025} and by shedding light on energy dissipation in convective regions under rapid tidal forcing \citep{Duguid2020,Vidal2020}. 
Observational tests remain pivotal. The distributions of orbital eccentricities and synchronization states provide constraints, but current samples are often limited by poorly characterized stellar properties and unknown inclinations. GAIA,  with precise parallaxes and orbital solutions, offers
unprecedented opportunities to map large populations of binary systems, allowing to test tidal theories across a wide range of stellar masses and evolutionary stages \citep{Dewberry2025}. Combining GAIA and seismic data from KEPLER, TESS and PLATO allows detailed probing of internal rotation and mode excitation, potentially revealing the impact of dynamical tides and resonance locking on orbital evolution \citep[e.g.][]{Burkart2012,Guo2022}. 
Heartbeat stars, with their tidally induced variability and oscillations, serve as unique laboratories for testing tidal dissipation and orbital evolution models \citep[e.g.][]{Fuller2017,Li2024}. 

In summary, while the last decade has seen remarkable progress in understanding binary interactions, fundamental uncertainties remain. Bridging these gaps will require a combination of advanced simulations, improved physical prescriptions in evolutionary models, and systematic observational tests, supported by next generation facilities. These efforts are crucial for developing a predictive theory of binary evolution that connects small-scale processes to the rich phenomenology observed across stellar populations.

\section*{Acknowledgments}
The binary evolution calculations used to generate some of the figures were computed with the BINSTAR code \citep{Siess2013}. The simulations shown in Fig.~\ref{fig:sph} have been performed with the smoothed particle hydrodynamics code PHANTOM \citep{Price2018,Siess2022}. LS is FNRS research director. 



\bibliographystyle{elsarticle-harv} 
\bibliography{reference}






\end{document}